\documentclass[11pt,a4paper]{article}
\pdfoutput=1
\usepackage{jheppub}

\usepackage{graphicx}

\usepackage{amssymb}
\usepackage{amsmath}
\usepackage{mathrsfs}
\usepackage{dsfont}

\usepackage{color}

\usepackage{vmargin}
\setmargins  {2.5cm}{1.5cm}% % linker & oberer Rand
             {15cm}{22.5cm}%   % Textbreite und -hoehe
             {28pt}{15pt}%   % Kopfzeilenhoehe und -abstand
             {0pt}{40pt}%    % \footheight (egal) und Fusszeilenabstand

\newcommand{\Eq}[1]{Eq.~(\ref{#1})}

\newcommand{\fig}[1]{Fig.~\ref{#1}}
\newcommand{\Fig}[1]{Fig.~\ref{#1}}

\newcommand{\tab}[1]{Table~\ref{#1}}

\title{Four-Flavour Leading-Order Hadronic Contribution To The Muon Anomalous Magnetic Moment}

\collaborationImg{\includegraphics[height=2cm]{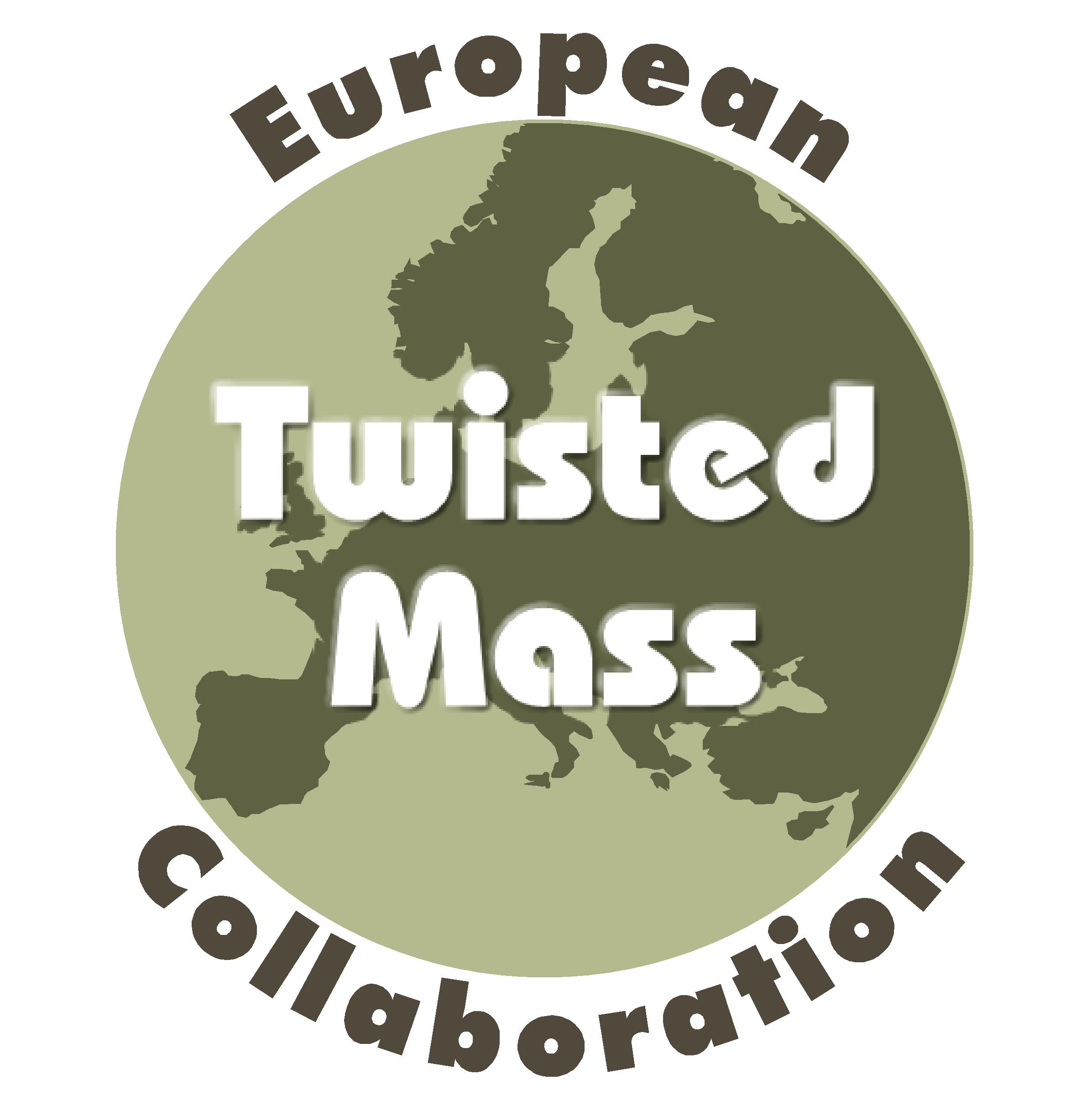}}

\author[a]{Florian Burger,}
\author[b]{Xu Feng,}
\author[a]{Grit Hotzel,}
\author[c,d]{Karl Jansen,}
\author[e]{Marcus Petschlies,}
\author[f]{Dru~B.~Renner}

\affiliation[a]{Humboldt-Universit\"at zu Berlin, Institut f\"ur Physik, Newtonstr. 15, D-12489 Berlin, Germany }
\affiliation[b]{High Energy Accelerator Research Organization (KEK), Tsukuba 305-0801, Japan}
\affiliation[c]{NIC, DESY, Platanenallee 6, D-15738 Zeuthen, Germany }
\affiliation[d]{Department of Physics, University of Cyprus, P.O.Box 20537, 1678 Nicosia, Cyprus}
\affiliation[e]{The Cyprus Institute, P.O.Box 27456, 1645 Nicosia, Cyprus}
\affiliation[f]{Jefferson Lab, 12000 Jefferson Avenue, Newport News, VA 23606, USA}

\emailAdd{grit.hotzel@physik.hu-berlin.de}

\abstract{
We present a four-flavour lattice calculation of the leading-order hadronic vacuum 
polarisation contribution to the anomalous magnetic moment of 
the muon,
$a_\mathrm{\mu}^{\rm hvp}$, arising from quark-connected Feynman graphs. 
It is based on ensembles featuring $N_f=2+1+1$ dynamical twisted mass
fermions generated by the European Twisted Mass
Collaboration (ETMC). Several light quark masses are used in order to yield a controlled extrapolation to the physical pion
mass. We employ three lattice spacings to examine lattice artefacts and several different volumes to
check for finite-size effects. Incorporating the complete first two generations of quarks allows for a direct comparison
with
phenomenological determinations of $a_\mathrm{\mu}^{\rm hvp}$. Our final result including an estimate of the systematic
uncertainty $$a_{\mathrm{\mu}}^{\rm hvp} = 6.74(21)(18) \cdot 10^{-8}$$ shows a
good overall agreement with 
these computations.
}

\arxivnumber{1308.4327}

\begin{document}
\maketitle

%\newpage

\section{Introduction}
\label{sec:intro}

The detection of a scalar boson at the LHC leaves us, in case that 
it turns out to be the standard model Higgs boson,  
with a somewhat puzzling situation. On the one hand, the standard model seems 
then to be complete. On the other hand, e.g. the lack of our understanding
of dark matter, the amount of CP violation, and the generation of baryon asymmetry in the universe
strongly suggest that new physics must exist. It is therefore a most important
task, to find indeed signs of this new physics beyond the standard model by looking 
at promising physical observables to detect its manifestations. 

The anomalous magnetic moment of the muon,
$a_{\mathrm{\mu}}$,
can be considered as such an important benchmark quantity to test
the standard model of particle interactions, or,
alternatively to detect new physics beyond the
standard model.
It can be
measured experimentally very precisely
\cite{Bennett:2006fi,Roberts:2010cj} and
can likewise be computed precisely in the
standard model, see e.g.~the review~\cite{Jegerlehner:2009ry}. A comparison between the experimental
result for $a_{\mathrm{\mu}}$ and a standard model
calculation reveals a discrepancy of about
three $\sigma$
which is persistent over many years
now and has been confirmed by computations of a number of groups.
The interesting question is then, whether this discrepancy originates
from some undetected effect in the experimental or theoretical
determination of $a_{\mathrm{\mu}}$, or, somewhat more excitingly,
whether it points
to physics beyond the standard model.

A key ingredient in the calculation
of $a_\mathrm{\mu}$ is
the leading order hadronic vacuum polarisation contribution,
$a_\mathrm{\mu}^{\rm hvp}$, which presently is the largest source
of uncertainty of the theoretical computation of
$a_\mathrm{\mu}$.
Since $a_\mathrm{\mu}^{\rm hvp}$ is of intrinsically
non-perturbative nature, evidently a lattice QCD computation for this
quantity is highly desirable. In fact, in a recent study, using
only two flavours of mass degenerate quarks, it could be demonstrated
that by employing an improved method to compute $a_\mathrm{\mu}^{\rm hvp}$
on the lattice, an unprecedented precision for
a lattice calculation of $a_\mathrm{\mu}^{\rm hvp}$
could be reached \cite{Feng:2011zk,Renner:2012fa}.

In this paper, we want to report on a calculation of $a_\mathrm{\mu}^{\rm hvp}$
with four flavours of quarks, using besides mass degenerate up and down quarks
also the strange and the charm quarks. Having such a setup is important for
two reasons. First, since the contribution of the
bottom quark is so small that neither experimental nor theoretical
determinations are currently sensitive to it, the standard model calculation is sensitive to four
flavours and disentangling the flavour contributions, for e.g. only the light
flavours, is afflicted with ambiguities. Using four flavours is
therefore cleaner and allows for a direct and unambiguous comparison
to phenomenological determinations.

The second reason is that the charm quark contribution computed in 
perturbation theory~\cite{Bodenstein:2011qy} is  $a_\mathrm{\mu, c}^{\rm hvp}
=1.44(1)\times10^{-9}$.
Hence, the charm quark contribution has approximately the same size as
the light-by-light contribution \cite{Prades:2009tw} and the electroweak 
contributions~\cite{Jegerlehner:2009ry} which are larger than the current experimental and theoretical uncertainties. Including the charm
quark contribution in a lattice calculation
is therefore important if values of $a_\mathrm{\mu}^{\rm hvp}$
are to be evaluated with a precision aiming to match the experimental errors.

For our computation of $a_\mathrm{\mu}^{\rm hvp}$ we will follow closely
the strategy of Refs.~\cite{Feng:2011zk, Renner:2012fa} applying an improved lattice definition
of $a_\mathrm{\mu}^{\rm hvp}$. As we will see below, also for the four
flavour calculation discussed here, this method
allows for a controlled extrapolation
to the physical point leading to an accurate determination
of $a_\mathrm{\mu}^{\rm hvp}$.

For our calculations
we employ configurations generated for four flavours by the
European Twisted Mass Collaboration (ETMC) \cite{Baron:2010bv,Baron:2010th}.
These sets
of configurations are obtained at different values of the lattice spacing  
and several volumes,
thus allowing us to estimate discretisation and finite size
effects
as systematic uncertainties in our lattice calculation.
In addition, at each value of the lattice spacing configurations exist
at several values of the pion mass, ranging from
$230\,\mathrm{MeV} \lesssim m_\pi \lesssim 450\,\mathrm{MeV}$ thus
allowing an extrapolation to the
physical point.

In this paper, we will give a detailed discussion about the 
extraction of $a_\mathrm{\mu}^{\rm hvp}$, providing our fitting strategy 
for the vacuum polarisation functions, and also address 
Pad\'e approximants to analyse the vacuum polarisation function as suggested in~\cite{Aubin:2012me}.

\section{Basic definitions}
\label{sec:def}
The key quantity for the determination of the leading-order hadronic contribution to the muon $g-2$ 
is the hadronic vacuum polarisation tensor
$\Pi_{\mu \nu}^{\mathrm{em}}(Q)$ with Euclidean momentum $Q$. It can be obtained from the correlator of two electromagnetic vector currents
\begin{equation}
J_{\mu}^{\mathrm{em}}(x) = 
\frac{2}{3} \overline{u}(x) \gamma_{\mu} u(x) - \frac{1}{3} \overline{d}(x) \gamma_{\mu} d(x) + \frac{2}{3}
\overline{c}(x) \gamma_{\mu} c(x) - \frac{1}{3} \overline{s}(x) \gamma_{\mu} s(x) 
\label{eq:current}
\end{equation}
in the following way
\begin{equation}
   \Pi_{\mu \nu}^{\rm em}(Q)= \int d^4 x \,e^{iQ\cdot(x-y)} \langle J_{\mu}^{\mathrm{em}}(x) J_{\nu}^{\mathrm{em}}(y)\rangle \; .
\label{eq:vptensor}
\end{equation}
The form of $J_{\mu}^{\mathrm{em}}$ given above anticipates that our simulations involve four 
dynamical quark flavours, up($u$), down($d$), strange($s$), and charm($c$).
Euclidean symmetry and the Ward-Takahashi identities require the vacuum polarisation tensor to be transverse, i.e.
\begin{equation}
   \Pi_{\mu \nu}^{\rm em}(Q)= (Q_{\mu} Q_{\nu} - Q^2 \delta_{\mu
\nu}) \Pi^{\rm em}(Q^2) \;.
\label{eq:transverse}
\end{equation}
$\Pi^{\rm em}(Q^2)$ is the hadronic vacuum polarisation function, for which the label ``em'' will be left out in the following to ease notation.  
Its renormalised variant 
\begin{equation}
\Pi_{\mathrm{R}}(Q^2)= \Pi(Q^2)- \Pi(0) 
\label{eq:pirenorm}
\end{equation}
appears in the
expression for the leading hadronic contribution to the anomalous magnetic moment of 
the muon in Euclidean space-time~\cite{deRafael:1993za, Blum:2002ii}
\begin{equation}
a_{\mathrm{\mu}}^{\mathrm{hvp}} = 
\alpha^2 \int_0^{\infty} \frac{d Q^2 }{Q^2} 
w\left( \frac{Q^2}{m_{\mathrm{\mu}}^2}\right) \Pi_{\mathrm{R}}(Q^2) \; ,
\label{eq:amudef}
\end{equation}
in which $\alpha$ denotes the fine-structure constant and $m_{\mathrm{\mu}}$ the muon mass.
Since the weight function $w\left( Q^2/ m_{\mathrm{\mu}}^2\right)$ is known from leading-order QED perturbation theory to have the form
\begin{equation}
 w(r) = \frac{64}{r^2 ( 1 + \sqrt{ 1 + 4/r } )^4 \sqrt{ 1 + 4/r }} \; , 
\label{eq:weightfun}
\end{equation}
the main task of the lattice calculation is the determination of
$\Pi_{\mathrm{R}}(Q^2)$ from the vector current correlator.

\section{Lattice calculation}
\label{sec:calc}

\subsection{Setup}
\label{sec:setup}
As mentioned before, the calculations are performed employing gauge field configurations generated by the 
ETMC with $N_f=2+1+1$
dynamical quark flavours. In the gauge sector the Iwasaki action~\cite{Iwasaki:1985we} is used. The up and down
quarks reside in a mass-degenerate flavour doublet $\chi_l =  \begin{pmatrix}
                                                                                          u \\ d
                                                                                         \end{pmatrix}$. 
They are described by the standard twisted mass lattice action~\cite{Frezzotti:2003ni}
\begin{equation}
 S_F[\chi_l, \overline{\chi}_l, U] =\sum_x \overline{\chi}_l(x) \left[ D_W +i \mu \gamma_5 \tau^3 \right] \chi_l(x)
\label{eq:Slight}
\end{equation}
where 
\begin{equation}
 D_W = \frac{1}{2} \gamma_{\mu} \left( \nabla_{\mu} + \nabla_{\mu}^* \right) - \frac{a}{2} \nabla_{\mu}^* \nabla_{\mu} + m_0
\end{equation}
is the Wilson Dirac operator with covariant forward and backward derivatives $\nabla_{\mu}$ and $\nabla_{\mu}^*$, respectively. The
value of $m_0$ has been
tuned to its critical value in order
to ensure automatic $\mathcal{O}(a)$ improvement of all physical observables. The bare twisted quark mass is
denoted by $\mu$ and $\tau^3$ is the third Pauli matrix acting in flavour space. 
For the heavy quarks $\chi_h =  \begin{pmatrix}
                                                                                          c \\ s
                                                                                         \end{pmatrix}$ the twisted mass action for a non-degenerate
fermion doublet~\cite{Frezzotti:2003xj}
\begin{equation}
  S_F[\chi_h, \overline{\chi}_h, U] =\sum_x \overline{\chi}_h(x) \left[ D_W +i \mu_{\sigma} \gamma_5 \tau^1 + \mu_{\delta}
\tau^3 \right] \chi_h(x)
\label{eq:Sheavy}
 \end{equation}
has been implemented with the same value for $m_0$. $\tau^1$ is the first Pauli matrix, $\mu_{\delta}$ denotes the mass splitting between the
charm and the strange quarks, and $\mu_{\sigma}$ the bare twisted mass with a twist in flavour space orthogonal to the one in the light sector.

Details of the analysed ensembles are given in \tab{tab:ensembles}. 
They involve different volumes, three different lattice spacings, and up to
five pion masses per lattice spacing such that finite-size effects, 
lattice artefacts, and the chiral extrapolation can be studied.

\begin{table}[htb]
\begin{center}
\begin{tabular}{|c | c c c c c|}
\hline
 & & & & &\vspace{-0.40cm} \\
Ensemble & $\beta$ & $a[{\rm fm}]$ & $\left(\frac{L}{a}\right)^3 \times  \frac{T}{a}$ & $m_{PS}$[MeV] &$ L$[fm] \\
 & & & & &\vspace{-0.40cm} \\
\hline \hline
& & & & &\vspace{-0.40cm} \\
D15.48 & $2.10$ & $0.061$ & $48^3 \times 96$ & 227 & 2.9    \\
D30.48 & $2.10$ & $0.061$ & $48^3 \times 96$ & 318 & 2.9    \\
D45.32sc& $2.10$ & $0.061$ & $32^3 \times 64$ & 387 & 1.9    \\
% & & & & \vspace{-0.40cm} \\
%
\hline
 & & & & & \vspace{-0.40cm} \\
B25.32t& $1.95$ & $0.078$ & $32^3 \times 64$ & 274 &  2.5 \\
B35.32 & $1.95$ & $0.078$ & $32^3 \times 64$ & 319 &  2.5  \\
B35.48 & $1.95$ & $0.078$ & $48^3 \times 96$ & 314 &  3.7  \\
B55.32 & $1.95$ & $0.078$ & $32^3 \times 64$ & 393 &  2.5  \\
B75.32 & $1.95$ & $0.078$ & $32^3 \times 64$ & 456 &  2.5 \\
B85.24 & $1.95$ & $0.078$ & $24^3 \times 48$ & 491 &  1.9 \\

 & & & & &\vspace{-0.40cm} \\

\hline
 & & & & &\vspace{-0.40cm} \\

A30.32 & $1.90$ & $0.086$ & $32^3 \times 64$ & 283 & 2.8 \\
A40.32 & $1.90$ & $0.086$ & $32^3 \times 64$ & 323 & 2.8 \\
A50.32 & $1.90$ & $0.086$ & $32^3 \times 64$ & 361 & 2.8 \\
\hline
\end{tabular}
\caption{ Parameters of the $N_f = 2+1+1$ flavour gauge field configurations that 
have been analysed in this work. $\beta$ denotes the gauge coupling, $a$ the lattice spacing, 
$\left(\frac{L}{a}\right)^3 \times  \frac{T}{a}$ the space-time volume, and $m_{PS}$ is the, unphysical, value of the pion mass. 
The values for $m_{PS}$ have been determined
in~\protect\cite{Baron:2010bv}. $L$ is
the spatial extent of the lattices. The approximate lattice spacings given here are 
taken from a first analysis of the used gauge field configurations~\protect\cite{Baron:2011sf}.}     
\label{tab:ensembles}                      
\end{center}
\end{table}

As mentioned before, the calculation of $\Pi(Q^2)$ requires the evaluation of the
vector-vector correlation function in Eq.~(\ref{eq:vptensor}). On the
lattice, one can use the local vector current
$J_{\mu}^L =  \overline{\chi}(x) Q_\mathrm{el} \gamma_\mu \chi(x)$ in complete analogy to
the continuum with the charge matrix $Q_\mathrm{el} = \mathrm{diag}(2/3,-1/3)$, where $\chi$ stands either for the light
or the heavy doublet.
However, due to the lattice regularisation this current has to be renormalised multiplicatively and does not satisfy a
Ward-Takahashi identity. To avoid these issues
we instead use the conserved Noether current also known as point-split vector current at sink and source position
\begin{equation}
 J_{\mu}^C(x)  =  \frac{1}{2} \left( \overline{q}(x+\hat{\mu})(\mathds{1}+\gamma_{\mu}) U_{\mu}^{\dagger}(x) q(x)  -
\overline{q}(x)(\mathds{1}-\gamma_{\mu}) U_{\mu}(x) q(x+\hat{\mu}) \right)\;.
\label{eq:conscurrent}
\end{equation}
This current is defined for each flavour separately using the decomposition 
of the electromagnetic vector current of \Eq{eq:current} into the quark currents
$J^{\rm u}$, $J^{\rm d}$, $J^{\rm c}$, and $J^{\rm s}$
\begin{equation}
 J_{\mu}^{\mathrm{em}}(x) = 
\frac{2}{3} J^{\rm u}(x) - \frac{1}{3}J^{\rm d} (x)+ \frac{2}{3} J^{\rm c} (x)- \frac{1}{3}J^{\rm s} (x)\; .
\end{equation}
Since the Noether current is only conserved for actions diagonal in flavour space, 
we employ an action for the heavy valence quarks different from
the sea quark action given in Eq.~(\ref{eq:Sheavy}), namely the so-called Osterwalder-Seiler (OS) action~\cite{Osterwalder:1977pc,
Frezzotti:2004wz}
\begin{equation*}
 S_F[\chi_h, \overline{\chi}_h, U] = \sum_x \overline{\chi}_h(x) \left[D_W  + i \left(\begin{array}{cc}
                                                                                                \mu_c & 0 \\
                                                                                                 0 & -\mu_s
                                                                                               \end{array} \right)
  \gamma_5
 \right] \chi_h(x) \;.
\end{equation*}
How to achieve $\mathcal{O}(a)$ improvement in this situation is discussed in
Ref.~\cite{Frezzotti:2004wz}.
The bare twisted mass parameters for the valence strange and the charm quarks, $\mu_s$ and $\mu_c$, are tuned in such a way that the physical
values for $2 m_K^2 - m_{\rm PS}^2$ and the
D-meson mass, respectively, are reproduced. Here, $m_K$ denotes the kaon mass. This leads to the following values
\begin{center}
\begin{tabular}{|c|cc|}
 \hline
 $\beta$ & $a \mu_s$ & $a \mu_c$ \\
 \hline
 \hline
 1.90   & 0.01815(10) & 0.2360(10) \\
 1.95   & 0.01685(10) & 0.2150(20) \\
 2.10   & 0.014165(10)& 0.1755(20) \\
 \hline
\end{tabular} 
.
\end{center}

With our definition of the vacuum polarisation tensor 
\begin{equation}
 \Pi_{\mu\nu}(x,y) = \langle J_\mu^C(x) J_\nu^C(y) \rangle + a^{-3} \delta_{\mu\nu} \langle S_\nu(y) \rangle\,,
\end{equation}
involving the contact term
\begin{equation}
 S_\nu(y) =
\frac{1}{2}\left\{ \overline{q}(y+\hat{\nu}) (1 + \gamma_\nu) U^\dagger_\nu(y) q(y)
+ \overline{q}(y) (1 - \gamma_\nu) U_\nu(y) q(y+\hat{\nu}) \right\}\,,
\end{equation}
we can now rely on the Ward
identity
\begin{equation}
 \partial_{\mu}^* \Pi_{\mu\nu}(x,y) = 0
\end{equation}
being valid at non-zero lattice spacings $a$.
%, where $\hat{Q}_{\mu} = \frac{2}{a} \sin{\frac{a q_{\mu}}{2}}$ 
%is the standard lattice momentum.
Hence, the vacuum polarisation tensor on the lattice satisfies in momentum space
\begin{equation}
   \Pi_{\mu \nu}(Q)= (\hat{Q}_{\mu} \hat{Q}_{\nu} - \hat{Q}^2 \delta_{\mu
\nu}) \Pi(Q^2) + \mathcal{O}(a) \;.
\label{eq:lattransverse}
\end{equation}
The reason for the $ \mathcal{O}(a)$ lattice artefacts is the lack of full Euclidean 
symmetry due to the lattice discretisation.
Note that nevertheless, as discussed below, the vacuum polarisation function 
can be made $ \mathcal{O}(a)$-improved.

\subsection{Vacuum polarisation}
\label{sec:vacpol}
Plugging \Eq{eq:current} into \Eq{eq:vptensor} and noting that there are no mixed-flavour contributions for the quark-connected Feynman diagrams, we
can obtain the total vacuum polarisation tensor by adding the vacuum polarisation tensors of the single-flavour contributions including the
appropriate charge factors
\begin{equation}
  \Pi_{\mu \nu}^{\rm tot}(Q) = \frac{5}{9}  \Pi_{\mu \nu}^{\rm ud}(Q) + \frac{1}{9}  \Pi_{\mu \nu}^{\rm s}(Q) + \frac{4}{9}  \Pi_{\mu
\nu}^{\rm c}(Q) \; .
\end{equation}
Notice that we have defined the single-flavour contributions to 
$\Pi_{\mu \nu}^{\rm tot}(Q)$ excluding the factors of the squared charges. In
the following we will generically denote them by $\Pi_{\mu \nu}(Q)$ and the respective 
vacuum polarisation function by $\Pi(Q^2)$. The factor of
$\Pi_{\mu \nu}^{\rm ud}(Q)$ is the sum of the squared charges of up and down quarks. 
Similarly, the total $a_{\mathrm{\mu}}^{\mathrm{hvp}}$ is
procured as a sum of the single-flavour contributions including the charge factors.

For constructing the vacuum polarisation function $\Pi(Q^2)$ from the vacuum 
polarisation tensor $\Pi_{\mu \nu}(Q)$ in \Eq{eq:lattransverse}, 
we define for
the single-flavour contributions
\begin{equation}
 \Pi(Q^2) = \Re{\left\lbrace \sum_{\mu, \nu, Q \in [Q]} \Pi_{\mu \nu}(Q) P_{\mu \nu}(Q) \left( \sum_{\mu, \nu, Q} P_{\mu \nu}(Q) P_{\mu \nu}(Q)
\right)^{-1}\right\rbrace} + \mathcal{O}(a^2)
\label{eq:piq2}
\end{equation}
as the real part of the projection onto the tensor $P_{\mu \nu}(Q) = \hat{Q}_{\mu} \hat{Q}_{\nu} - \hat{Q}^2 \delta_{\mu
\nu}$ with $Q^2 = \sum_{\mu} \hat{Q}^2_{\mu}$. 
The class $[Q]$ is defined by all momenta related by a Euclidean space-time symmetry transformation,
although these are not all exact symmetries at non-zero lattice spacing and finite volume. 
It has been checked that this does not lead to observable effects
in $\Pi(Q^2)$
in the low-momentum region, $Q^2 \le 2\,{\rm GeV}^2$, which is by far the most 
important for the determination of $a_{\mathrm{\mu}}^{\mathrm{hvp}}$.
In particular, the definition in \Eq{eq:piq2} involves summing over $Q$ and $-Q$ which is 
required for achieving $\mathcal{O}(a)$
improvement. 

Following a similar line of arguments as given in~\cite{Cichy:2013yea, Cichy:2013zea}, it can be demonstrated that also for
the case of the vacuum polarisation function defined as in Eq.~(\ref{eq:piq2}) short-distance singularities do not spoil automatic
$\mathcal{O}(a)$ improvement at maximal twist~\cite{Burger:2014}. Therefore, continuum extrapolations are performed with an $a^2$-term in
Sec.~\ref{sec:results}.
 
Due to the renormalisation of the vacuum polarisation function given in
\Eq{eq:pirenorm}, the calculation of $a_{\mu}^{\mathrm{hvp}}$ for each flavour involves both an interpolation of 
$\Pi(Q^2)$ in between discrete lattice momenta  as 
well as an extrapolation to zero momentum. Hence, the choice of
fit function is important, especially in the low-momentum region where the weight function in \Eq{eq:weightfun} is peaked, and 
its influence on the calculation of $a_{\mu}^{\mathrm{hvp}}$ has to be investigated.

Our standard fit of the low-momentum dependence of the vacuum polarisation function involves two different terms.
The first is inspired by vector meson dominance with 
$M$ vector meson mass poles 
\begin{equation}
  \Pi_{\mathrm{low}}(Q^2) = \sum_{i=1}^M \frac{f^2_i}{m^2_i + Q^2} + \sum_{j=0}^{N-1} a_j (Q^2)^{j} \; ,
\label{eq:pilow}
\end{equation}
whereas the second term parametrises remaining deviations in the low-momentum region which
extends up to a matching momentum $Q^2_\mathrm{match}$. Here, $m_i$ denotes the mass of the 
vector meson states and $f_i$
their decay constants. They are determined before fitting the vacuum polarisation from the same vector correlation functions
partially Fourier transformed in the spatial directions on the same bootstrap samples as will be described in the next section. Thus, the
$a_j$ are the only parameters fitted here. For the matching of low and high momentum
fit functions we have used $Q^2_\mathrm{match} = 2 \mathrm{~GeV}^2$.

The high-momentum fraction of $\Pi(Q^2)$ for $Q^2 > Q_\mathrm{match}^2$ is interpolated, 
inspired by perturbation theory,  
using a polynomial in $Q^2$ and a polynomial multiplied by a
logarithmic term 
\begin{equation}
  \Pi_{\mathrm{high}}(Q^2) = \log(Q^2) \sum_{k=0}^{B-1} b_k (Q^2)^{k}  + \sum_{l=0}^{C-1} c_l (Q^2)^{l} \; . 
\end{equation}
Here, the fit parameters are denoted $b_k$ and $c_l$.
The total vacuum polarisation function for each flavour is then obtained from
\begin{equation}
 \Pi(Q^2) = (1- \Theta(Q^2-Q^2_{\rm match}))\Pi_{\mathrm{low}}(Q^2) + \Theta(Q^2-Q^2_{\rm match}) \Pi_{\mathrm{high}}(Q^2) 
\end{equation}
with $\Theta(Q^2)$ denoting the Heaviside step function.

\begin{figure}[htb]
\centering
\includegraphics[width=0.9\textwidth]{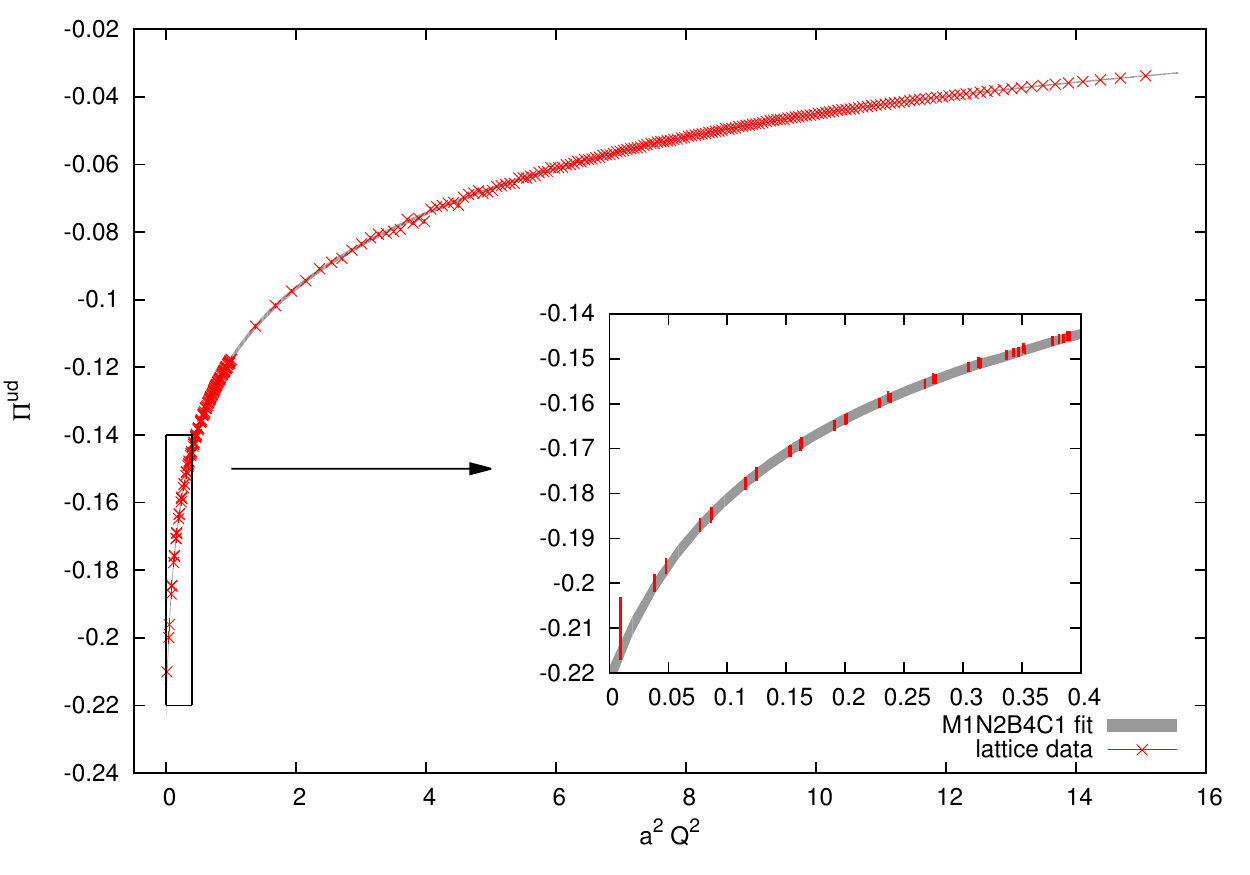}
\caption{Unrenormalised vacuum polarisation function of the light quarks, $\Pi^{\rm ud}(Q^2)$, for ensemble B25.32t (see
\protect\tab{tab:ensembles} for
details).} 
\label{fig:pilight}
\end{figure}

For the light and the strange quark contributions to $\Pi_{\mu \nu}^{\rm tot}(Q)$ 
our standard fit is characterised by $M=1$, $N=2$, $B=4$, and
$C=1$. An example fit for the vacuum polarisation for one of our lightest 
pion masses (B25.32t in \tab{tab:ensembles}) is given in
\Fig{fig:pilight}. Since the curvature of the vacuum
polarisation function for the charm quark contribution is much smaller, more parameters are needed in the high-momentum domain in order to ensure
smooth contact between the fit functions in the low and high momentum regions. 
Thus we employ for the charm quark contribution $M=1$, $N=2$, $B=3$, and $C=5$ in our standard
fit. The effect of varying the number of terms has been checked and will be reported when 
discussing the systematic uncertainties of our calculation.

To minimise finite-size effects we have excluded the points with the lowest 
lattice momentum in the vacuum polarisation fits which due to the big uncertainty of these points has only a very mild influence on the fits. This has
been suggested by a tree-level study showing that except for the lowest momentum 
point all other data points for different volumes fall on top of each
other.

\subsection{Vector meson mass fits}
\label{sec:vectormesons}
Since the $\rho$-meson mass will be needed in the chiral extrapolation as will be 
described below and since \Eq{eq:pilow} involves the masses
and decay constants of the vector meson states, we determine these basic quantities from the same
vector-vector correlators, i.e.~using the vector currents as interpolating operators for the vector mesons. Employing
the same bootstrap samples as in the vacuum polarisation fits, the uncertainties of the
determination of the vector meson properties can be correctly propagated to $a_{\mathrm{\mu}}^{\mathrm{hvp}}$. In fact, our way of chirally
extrapolating the data to the physical point benefits from a cancellation of the error of the $\rho$-meson mass achieved in this way.

In principle, the vector mesons should be
treated as resonances. However, due to angular momentum conservation the decay products, 
two pions in the case of the light quarks, can only be
produced if the kinematical condition $m_{\rm V} \ge 2 \sqrt{m_{\rm PS}^2+ \vec{p}^2}$ is 
satisfied with $\vec{p}\neq 0$. Since on the lattice with finite spatial
extent $L$, the momenta are quantised, the above condition becomes
\begin{equation}
 m_{\rm V} \ge 2 \sqrt{m_{\rm PS}^2+ \left(\frac{2 \pi}{L}\right)^2} \; .
\end{equation}
This condition is not fulfilled for all but one   
ensemble (D15.48 of \tab{tab:ensembles}) where 
the vector meson mass $m_{\rm V}$ and the energy of the 2-pion state with non-zero momentum become
consistent within errors. We nevertheless treat the lightest vector meson as a 
stable asymptotic state for all ensembles \footnote{We do not expect a significant 
effect of taking the resonance mass or the correlator mass in our analysis, since in the 2-flavour case a comparison
between~\cite{Feng:2010es} and~\cite{Feng:2011zk} gave consistent results.} 
and obtain the spectral information from the
large-time behaviour of the correlator projected to zero momentum in order to utilise
it in the vacuum polarisation fits, including also the
error propagation to $a_{\mu}^{\mathrm{hvp}}$. 

Following the analysis described in~\cite{Jansen:2009hr} we adopt
\begin{align}
 C(t)=& \sum_{\vec{x}} \sum_{k=1}^{3} \langle J_k^C(t, \vec{x}) J_k^C(0)\rangle \nonumber \\
 & \xrightarrow{t \to \infty} 3 m_V f_V^2 e^{- m_V
\frac{T}{2}} \cosh\left(\left(\frac{T}{2}-t\right)m_V\right) 
\end{align}
for our correlated fits to extract $m_V$ and $f_V$ from it. The factor three arises from the 
polarisation sum of the vector meson. Likewise fits
including
$M-1$ excited state contributions are performed with
\begin{equation}
 C(t; M) = \sum_{i=1}^{M} 3 m_i f_i^2 e^{- m_i
\frac{T}{2}} \cosh\left(\left(\frac{T}{2}- t\right)m_i\right)
\end{equation}
in an appropriate fit range. The statistical uncertainties of the fit parameters are 
estimated using the bootstrap method.

For single-state fits the initial timeslice for the fit should be large
enough such that the first excited state is sufficiently suppressed. 
The final fitting timeslice should be small enough to avoid the noisiest part of the
correlator which is obtained from simple point sources in our calculation. 
Taking those restrictions into account we have selected fixed time ranges
in physical units for the single-state fits by requiring the mean $\chi^2/dof$ for 
{\em all the ensembles} to be close to 1. To fit the $\rho$-meson
properties our standard fit range is $0.7\,\mathrm{fm} < t < 1.2\,\mathrm{fm}$. 
For the $\overline{s}s$-state we have chosen $0.9\,\mathrm{fm} < t <
1.4\,\mathrm{fm}$ and for the $J/\Psi$ fits $1.2\,\mathrm{fm} < t < 1.7\,\mathrm{fm}$. 
In the appendix in \tab{tab:paras_rho_ss} and
\tab{tab:paras_JPsi} we list the results for $m_V$ and $f_V$ obtained with 
single-state fits in our standard fit ranges. 
For one value of the lattice spacing the results for the $\rho$-meson masses are also depicted in \Fig{fig:rhomass}.
For fits with $M=2$ done to
check systematic effects of choosing a MNBC fit function, we have used 
$0.3\,\mathrm{fm} < t < 1.2\,\mathrm{fm}$ in the light sector,
$0.35\,\mathrm{fm} < t < 1.4\,\mathrm{fm}$ in the strange sector, and  
$0.4\,\mathrm{fm} < t < 1.7\,\mathrm{fm}$ in the charm sector. 
We include the systematic effect of choosing different ranges for fitting
the vector meson properties in our total error budget which will be discussed below.

\subsection{Chiral extrapolation of $a_\mu$}
As in~\cite{Feng:2011zk, Renner:2012fa} we use the modified definition
\begin{equation}
 a_{\overline{\mathrm{\mu}}}^{\mathrm{hvp}} = \alpha^2 \int_0^{\infty} \frac{d Q^2 }{Q^2} w\left( \frac{Q^2}{H^2}
\frac{H_{\mathrm{phys}}^2}{m_{\mathrm{\mu}}^2}\right) \Pi_{\mathrm{R}}(Q^2) 
\label{eq:redef}
\end{equation}
to determine $a_{\mathrm{\mu}}^{\mathrm{hvp}}$ with the same motivation as outlined
there. $H$ stands for some hadronic scale 
that can be determined at unphysical values of the pion
mass $m_{\rm PS}$. It is required to have a well-defined limit at the 
physical pion mass, $m_\pi$, denoted $H_{\rm phys}$ which has to be known from
experimental measurements for example. With such a choice, by definition 
$a_{\overline{\mathrm{\mu}}}^{\mathrm{hvp}} \rightarrow a_{\mathrm{\mu}}^{\mathrm{hvp}}$ when $H
\to H_{\rm phys}$.

\begin{figure}[ht]
 \centering
 \includegraphics[width=0.75\textwidth]{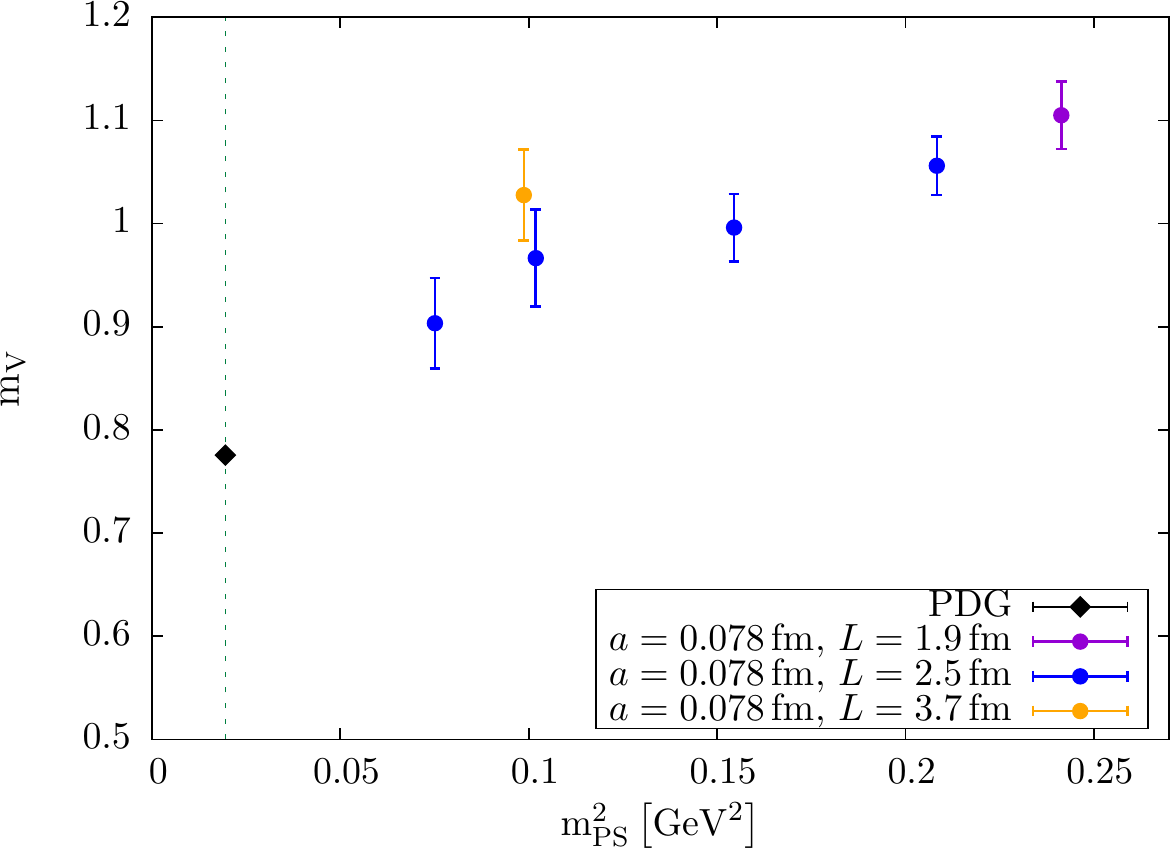}
 \caption{Mass of the $\rho$-vector meson as a function of the squared pion mass for the $\beta = 1.95$ ensembles (see
\protect\tab{tab:ensembles}).} 
 \label{fig:rhomass}
 \end{figure}

Inspired by the observation that the $\rho$-meson gives the
dominant contribution to $a_{\mathrm{\mu}}^{\mathrm{hvp}}$, we have chosen 
$H=m_V$ in the following. Here, $m_V$ denotes the $\rho$-meson mass for
unphysical values of the light quark masses. 
As noted in Ref.~\cite{Feng:2011zk} the $\rho$-meson mass determined from the time-dependent correlator
attains unphysically high values at unphysical values of the pion mass which cannot reliably be extrapolated to the physical point. This
can also be seen for our $N_f=2+1+1$ ensembles for one value of the lattice spacing in \Fig{fig:rhomass}.

Also in the four-flavour case the condition that the $\rho$-meson mass attains its physical value when the pion mass does,
seems to hold. Thus, as outlined in ~\cite{Feng:2011zk}, choosing $H=m_V$ amounts to an absorption of a 
large part of the pion mass dependence of 
$a_{\mathrm{\mu}}^{\mathrm{hvp}}$ in the light sector and to an effective 
redefinition of the muon mass on the lattice given by
\begin{equation}
 m_{\overline{\mu}} = m_{\mu}\cdot \frac{H}{H_{\rm phys}} \; .
\end{equation}
Since we want to use a consistent definition of the so modified muon mass 
for all single-flavour contributions to
$a_{\mathrm{\mu}}^{\mathrm{hvp}}$, we
also use the $\rho$-meson mass $m_V$ to redefine $a_{\mathrm{\mu}}^{\mathrm{hvp}}$ 
for the strange and the charm quarks.

It is interesting to note that in this way we introduce a non-linear 
pion mass dependence for the heavy flavour contributions. This is in contrast to 
the standard definition in \Eq{eq:amudef} where the pion mass dependence of $a_{\mathrm{\mu, s}}^{\mathrm{hvp}}$ and
$a_{\mathrm{\mu, c}}^{\mathrm{hvp}}$ is rather flat. 
However, since the heavy flavours constitute only a small fraction of the total $a_{\mathrm{\mu}}^{\mathrm{hvp}}$, 
the net effect of the modified definition of $a_{\mathrm{\mu}}^{\mathrm{hvp}}$ on the lattice 
is that the total $a_{\overline{\mathrm{\mu}}}^{\mathrm{hvp}}$ still exhibits a flat pion mass dependence.

\section{Results}
\label{sec:results}

\subsection{The light quark contribution, $a_{\mathrm{\mu, ud}}^{\mathrm{hvp}}$}
\begin{figure}[htb]
\begin{minipage}{0.49\textwidth}
\includegraphics[width=\textwidth]{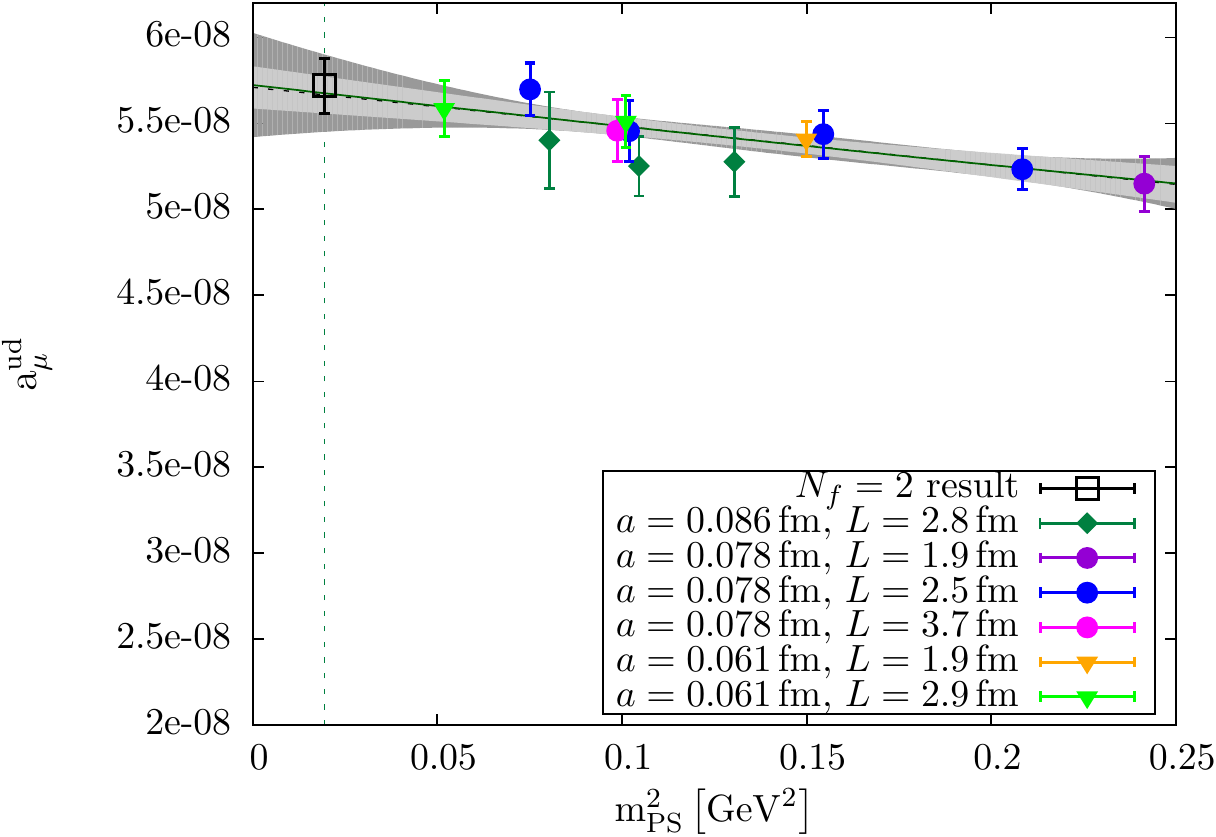}
\caption{Light-quark contribution to $a_{\mathrm{\mu}}^{\mathrm{hvp}}$ on $N_f=2+1+1$ sea.} 
\label{fig:amulight}
\end{minipage}
\hspace{0.02\textwidth}
\begin{minipage}{0.49\textwidth}
\includegraphics[width=\textwidth]{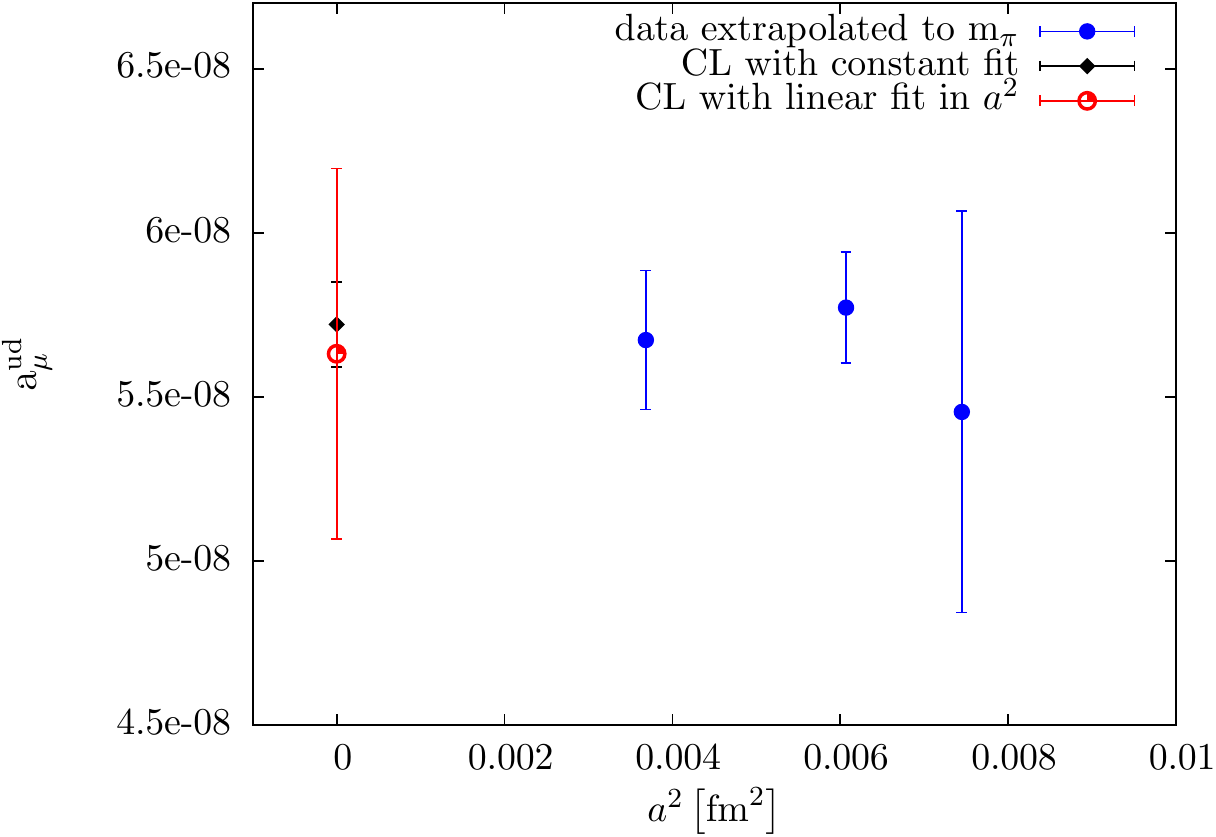}
\caption{Continuum extrapolation of $a_{\mathrm{\mu}}^{\mathrm{light}}$.} 
\label{fig:amulight_cl} 
\end{minipage}
\end{figure}

Considering only the light currents, for which the sea quark action is 
identical to the valence action, provides the contribution of the up and down quarks to the total
$a_{\mathrm{\mu}}^{\mathrm{hvp}}$. The pion mass dependence of this light 
quark portion is shown in \Fig{fig:amulight}. Here we utilise our lattice redefinition
$a_{\mathrm{\overline{\mu}, ud}}^{\mathrm{hvp}}$, i.e. \Eq{eq:redef} with $H=m_V$. 
This allows for a linear
extrapolation in the squared pion mass, $m_{\rm PS}^2$ (broken black line 
with light grey error-band). A quadratic fit to the data (green
solid line with dark grey error-band) gives almost the same value with a bigger uncertainty at the physical point.

The value at the physical point obtained by the linear fit can be compared 
to the value obtained with only two dynamical quark
flavours from our earlier lattice QCD analysis~\cite{Feng:2011zk}
\begin{eqnarray}
a_{\mathrm{\mu},\mathrm{ud}}^{\mathrm{hvp}} & = & 5.67(11)\cdot 10^{-8}\;\;\; (N_f=2+1+1)  \nonumber\\
a_{\mathrm{\mu},\mathrm{ud}}^{\mathrm{hvp}} & = & 5.72(16)\cdot 10^{-8}\;\;\; (N_f=2)
\label{eq: lighcontribution}
\end{eqnarray}
yielding fully compatible results. The difference between the error of the 
two results is that the $N_f = 2+1+1$ uncertainty given above is only of
statistical nature whereas the $N_f = 2$ value involves an estimate of systematic effects. 
The above result has been
obtained by fitting all data from the ensembles listed in \tab{tab:ensembles} 
simultaneously as the present quality of our data does not allow to
discriminate lattice artefacts in the light sector. This is shown in 
\Fig{fig:amulight_cl}. Here, we have first extrapolated
$a_{\mathrm{\overline{\mu}, ud}}^{\mathrm{hvp}}$ linearly
to the physical point fixing the value of the lattice spacing. 
The figure shows that all chirally extrapolated values agree 
within the errorbars. We can therefore use a constant extrapolation to
zero
lattice spacing giving $a_{\mathrm{\mu, ud}}^{\mathrm{hvp}}=5.72(13)\cdot 10^{-8}$ 
which is compatible with the result quoted in \Eq{eq: lighcontribution}.
We also performed a 
combined fit in $m_{\rm PS}^2$ and $a^2$ to all the data in \Fig{fig:amulight} yielding a
coefficient of the $a^2$-term compatible with zero.
Hence, for the present level of precision of our data, $a_{\mathrm{\mu},\mathrm{ud}}^{\mathrm{hvp}}$ 
does not show any significant lattice spacing artefacts.

\subsection{The three-flavour contribution, $a_{\mathrm{\mu, uds}}^{\rm hvp}$}
\begin{figure}[htb]
\begin{minipage}{0.49 \textwidth}
\includegraphics[width=\textwidth]{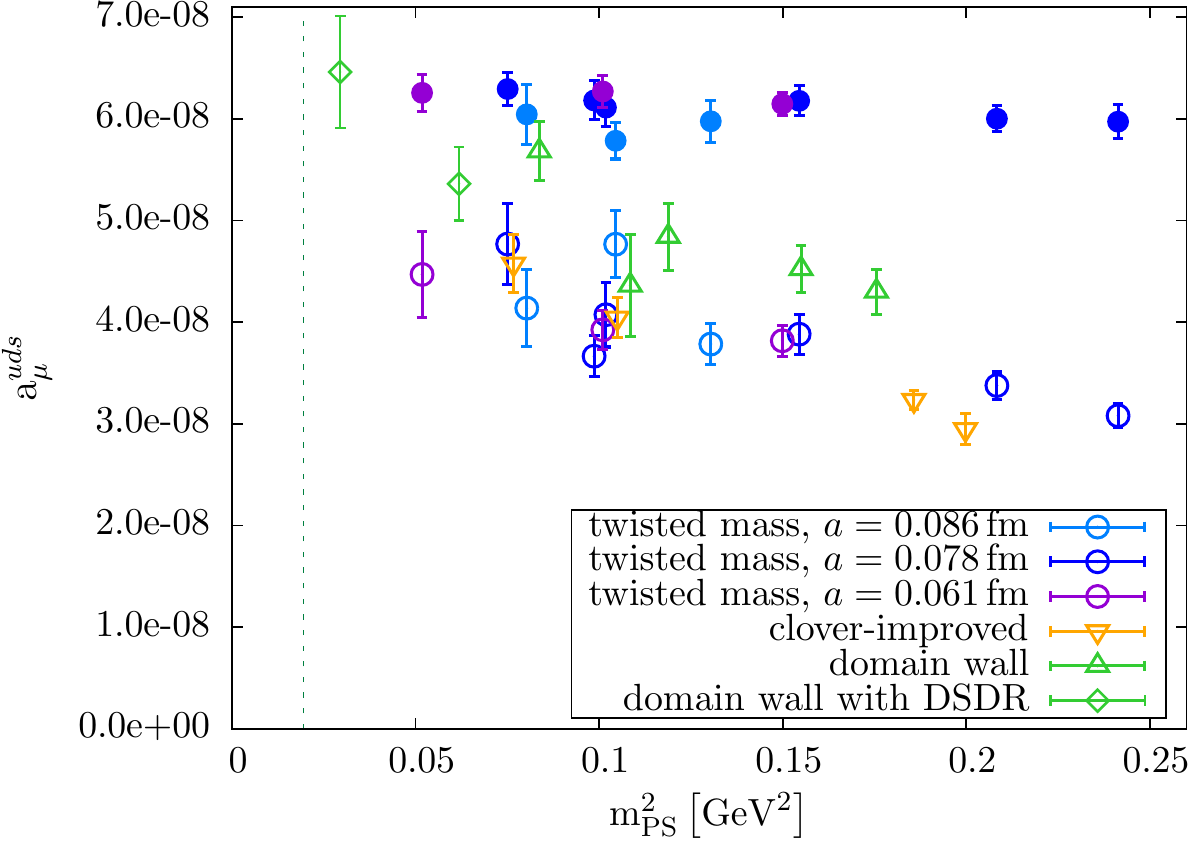}
\caption{Comparison of three-flavour contribution to $a_{\mathrm{\mu}}^{\mathrm{hvp}}$ obtained with different fermion
actions.} 
\label{fig:amunf21_comp}
\end{minipage}
\hspace{0.02\textwidth}
\begin{minipage}{0.49\textwidth}
\includegraphics[width=\textwidth]{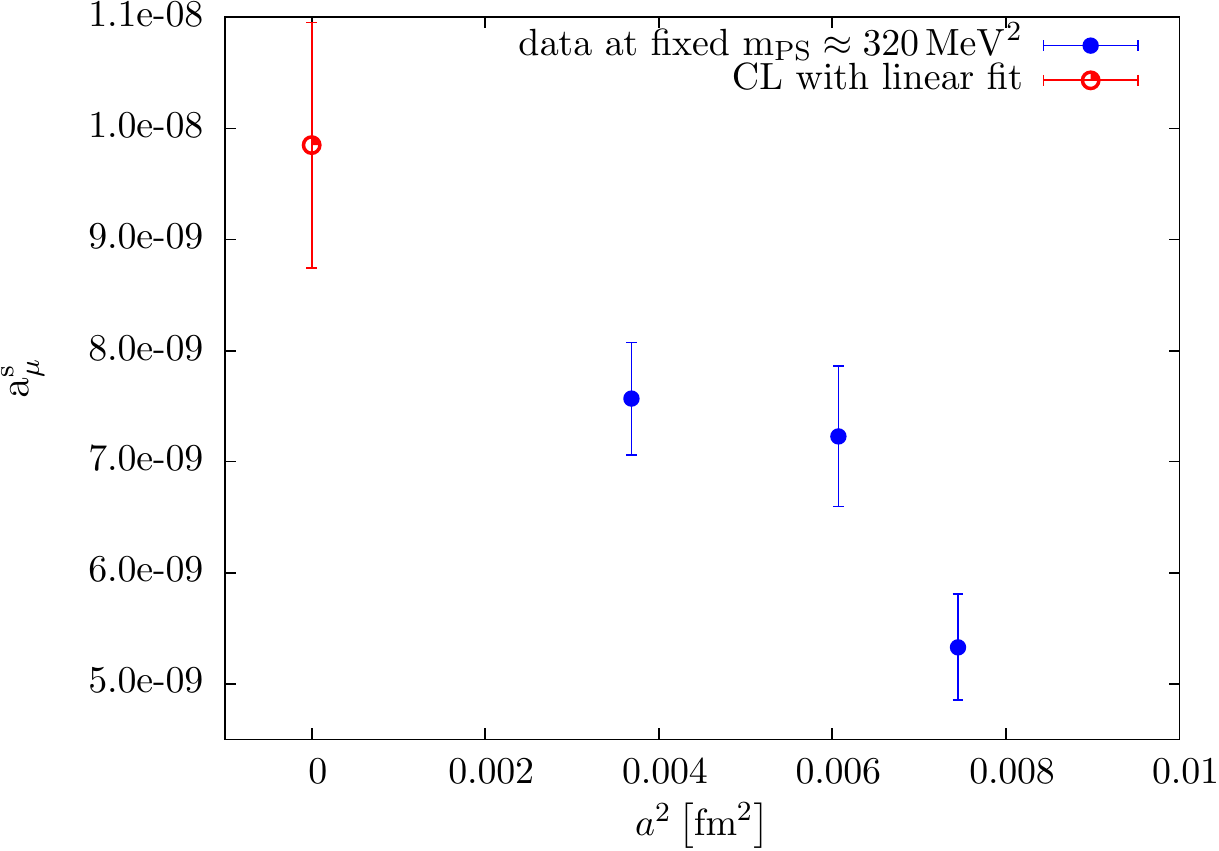}
\caption{Continuum extrapolation of $a_{\mathrm{\mu, s}}^{\rm hvp}$ at $m_{\rm PS}\approx320\,{\rm MeV}$.}
\label{fig:amustrange_cl}
\end{minipage}
\end{figure}

For the three-flavour contribution we use again $H=m_V$ as the hadronic scale 
in order to have a consistent redefinition of the muon mass on the
lattice. It turns out that this leads to larger statistical uncertainties for the 
strange quark contribution than employing the standard definition of 
$a_{\mathrm{\mu, s}}^{\rm hvp}$.  In addition, 
the 
dependence of $a_{\mathrm{\mu, s}}^{\rm hvp }$ on the squared pion mass appears to be 
non-linear. 
However, since the light quark contribution constitutes by far
the largest part of $a_{\mathrm{\mu}}^{\rm hvp}$, 
we still obtain a mild pion mass dependence for $a_{\mathrm{\mu, uds}}^{\rm hvp}$ as can be seen
when
looking at the twisted mass points (upper set of data points with filled symbols) in \Fig{fig:amunf21_comp}. 

In this figure we also include data obtained with different fermion
actions naturally possessing differing cut-off effects from the literature. The orange downward triangles
are
from~\cite{DellaMorte:2011aa} 
using clover-improved Wilson
fermions with two dynamical light and a quenched strange quark. 
The green upward triangles and diamonds have been computed with $N_f=2+1$ dynamical domain wall
fermions~\cite{Boyle:2011hu}. These results from the other groups employ the standard 
definition for $a_{\mathrm{\mu}}^{\mathrm{hvp}}$ given in \Eq{eq:amudef}. 
We therefore add also twisted mass points obtained with the standard definition 
(lower set of twisted mass points 
with open symbols). We find an overall agreement between the different lattice 
determinations for the raw data of $a_{\mathrm{\mu, uds}}^{\rm hvp}$ when 
the standard definition is used. Besides the aforementioned cut-off effects, slight differences can also originate from 
utilising different conditions to determine the strange quark mass used for computing the 
strange quark contribution to $a_{\mathrm{\mu, uds}}^{\rm hvp}$. 
It is clear from \Fig{fig:amunf21_comp} that
the improved definition of $a_{\mathrm{\mu}}^{\mathrm{hvp}}$ leads to 
a smooth and linear extrapolation to the physical point. In contrast, 
the standard definition of $a_{\mathrm{\mu}}^{\mathrm{hvp}}$
needs a more complicated extrapolation resulting in a larger 
uncertainty. 

Since the strange quark is heavier than the light quarks, the size of 
lattice artefacts can be significantly enhanced for the strange quark 
contribution. In fact, if we 
take lattice artefacts into account performing a combined fit in 
$m_{\rm PS}^2$ and $a^2$, we find a non-zero value for the coefficient of the 
$a^2$-term. The
presence of lattice artefacts can also be seen when looking at the lattice spacing dependence of $a_{\mathrm{\overline{\mu}, s}}^{\rm hvp}$ at a fixed
pion mass of about $320\,{\rm MeV}$ as shown in \Fig{fig:amustrange_cl}.
Here, we clearly see that the limit 
$a \to 0$ can no longer be obtained by a constant extrapolation.

\begin{figure}[htb]
\centering
\includegraphics[width=0.7\textwidth]{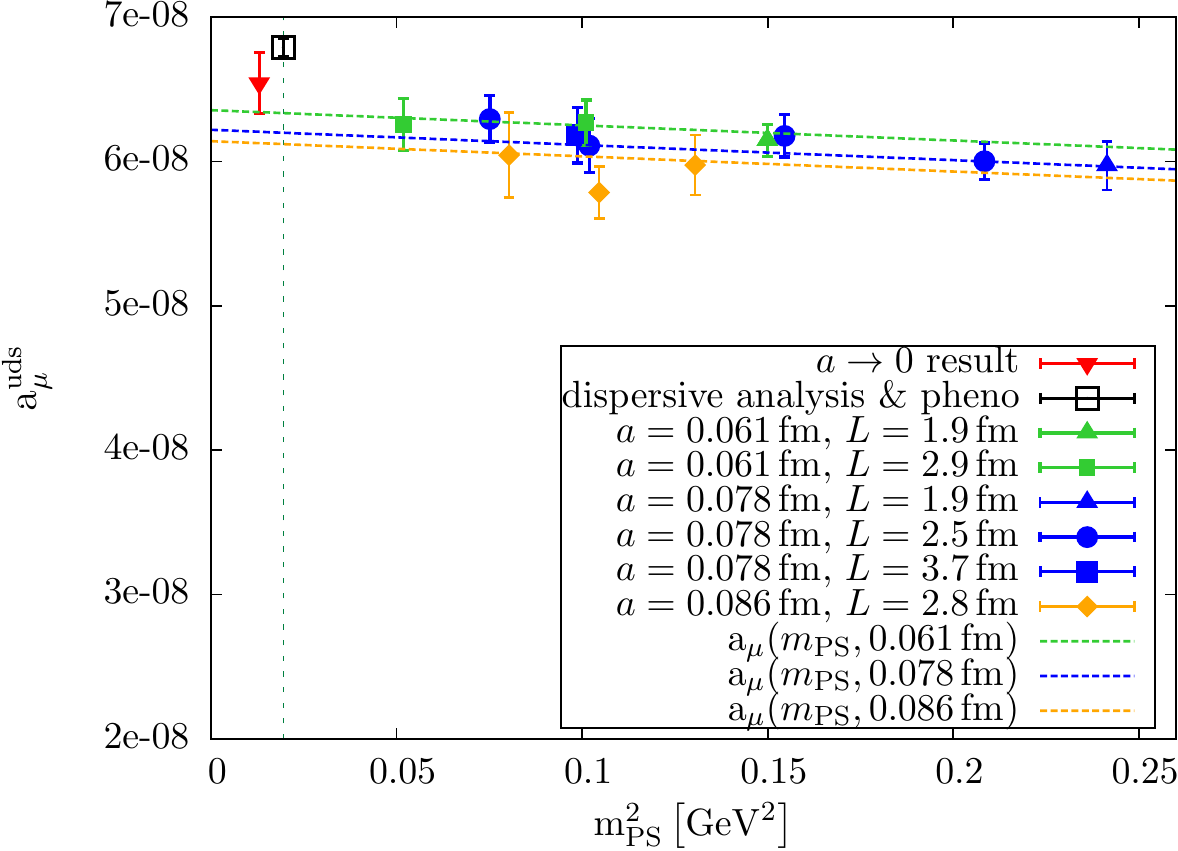}
\caption{Three-flavour contribution to $a_{\mathrm{\mu}}^{\mathrm{hvp}}$. The 
phenomenological value is extracted from~\protect\cite{Jegerlehner:2008zz}
assuming quark-hadron duality.} 
\label{fig:amunf21}
\end{figure}

As a result of this observation it is clear that we have to take an $a^2$-term into account when we discuss the continuum and chiral
extrapolation of $a_{\mathrm{\mu, uds}}^{\rm hvp}$. We therefore 
use the following fit function to obtain
$a_{\mathrm{\mu, uds}}^{\rm hvp}$
at the physical point
\begin{equation}
 a_{\mathrm{\mu}}(m_{\rm PS}, a) = A + B~m_{PS}^2 + C~a^2
\label{eq:fit}
\end{equation}
with $A, B, C$ denoting the free parameters of the fit. 
The value resulting from this fit at $m_{\rm PS} = m_{\pi}$ and zero lattice spacing is represented in
\Fig{fig:amunf21}
by the red triangle slightly displaced from the physical point in order to facilitate the 
comparison with the phenomenological result.  The dashed
lines represent $a_{\mathrm{\mu, uds}}^{\rm hvp}(m_{\rm PS}, a)$ at fixed values of the 
lattice spacing as indicated by the colours.
We have 
checked in addition that performing chiral and continuum limit extrapolations 
independently yields a compatible result with the one obtained from the combined fit.

In order to compare our three-flavour value to a
result from a dispersive analysis, we need to disentangle the quark flavours. 
To this end,  
we have to reweight the total $a_{\mathrm{\mu}}^{\mathrm{hvp}}$. There are 
different possibilities to carry out such a reweighting. We have decided to reweight the values
given in~\cite{Jegerlehner:2008zz} by the charges of the active flavours. 
This approach is based on the assumption of quark-hadron duality.
Given the ambiguity of such an approach 
we indicate by the abbreviation ``pheno'' that a certain phenomenological analysis has been 
employed. Comparing our lattice result with this phenomenological extraction method leads to
\begin{eqnarray}
a_{\mathrm{\mu, uds}}^{\rm hvp} & = & 6.55(21)\cdot 10^{-8}\;\;\; (N_f=2+1+1)  \nonumber\\
a_{\mathrm{\mu, uds}}^{\rm hvp} & = & 6.79(05)\cdot 10^{-8}\;\;\; (\mathrm{pheno})
\label{eq:nf21contribution}
\end{eqnarray}
where we find, at least within the errors, an agreement. 
Given the fact that our phenomenological value is certainly afflicted 
by an unkown systematic error, we consider it reassuring that our lattice QCD analysis can reproduce the 
phenomenological value at this level of accuracy. 
As mentioned in the
introduction, this ambiguity in the comparison of lattice results and those utilising the dispersion relation can be
removed by the inclusion of the charm quark in the calculation which we want to report on next.

\subsection{The four-flavour contribution, $a_{\mathrm{\mu}}^{\rm hvp}$}
\begin{figure}[htb]
\centering
\includegraphics[width=0.7\textwidth]{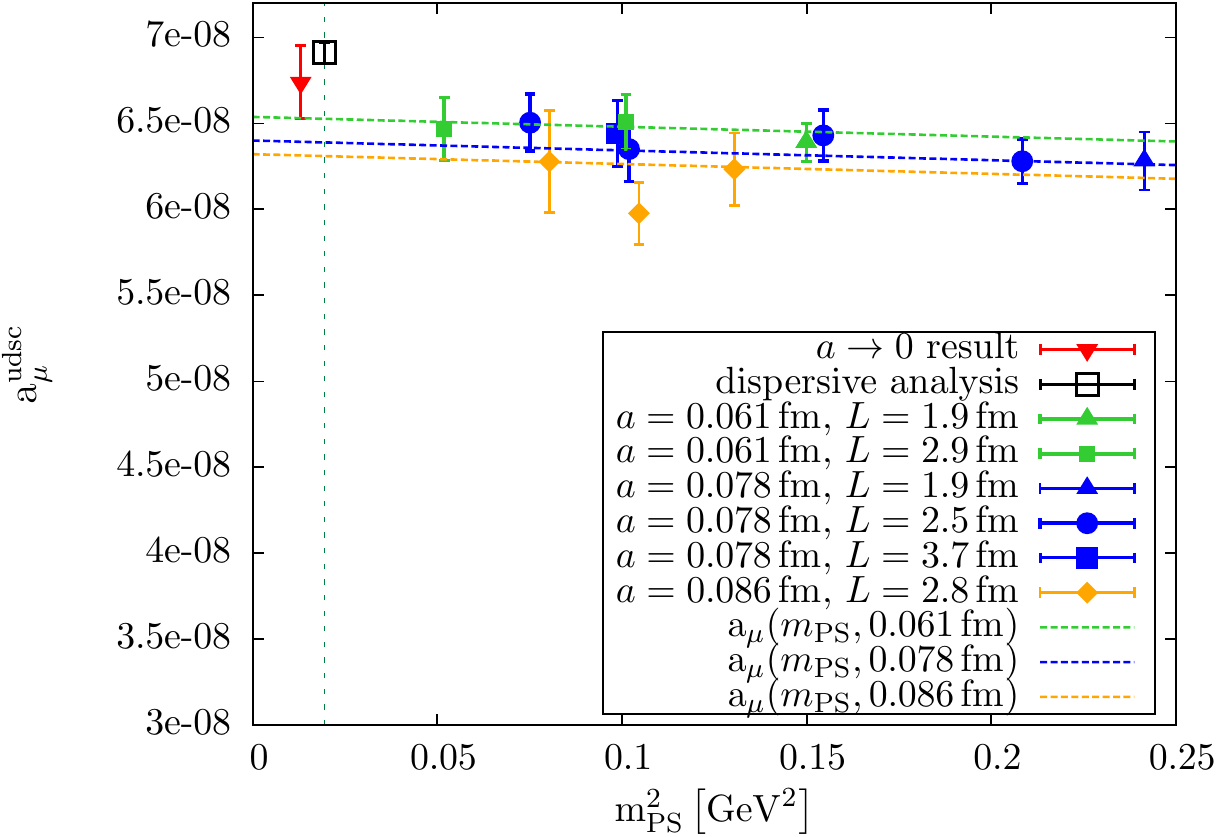}
\caption{$N_f=2+1+1$ result for $a_{\mathrm{\mu}}^{\rm hvp}$.} 
\label{fig:amutot}
\end{figure}

Adding the charm quark contribution according to \Eq{eq:redef} again using $H=m_V$  
we are able to
directly compare to experimental values and those from different dispersive analyses. 
Since the charm quark is even heavier than the strange quark, we 
again use a combined fit involving an $m_{\rm PS}^2$- and
an $a^2$-term of the form stated in \Eq{eq:fit}. In this way, we arrive at the picture shown in 
\Fig{fig:amutot}. Here, our result obtained in the continuum limit
and at a physical value of the pion mass, represented by the red triangle,
can now be unambiguously confronted with the corresponding one from 
a dispersive analysis~\cite{Jegerlehner:2008zz}:
\begin{eqnarray}
a_{\mathrm{\mu}}^{\rm hvp} & = & 6.74(21)\cdot 10^{-8}\;\;\; (N_f=2+1+1)  \nonumber\\
a_{\mathrm{\mu}}^{\rm hvp} & = & 6.91(05)\cdot 10^{-8}\;\;\; (\mathrm{dispersive}\; \mathrm{analysis}) \; .
\label{eq:amutot}
\end{eqnarray}
Comparing the value of the total $a_{\mathrm{\mu}}^{\rm hvp}$ 
now a convincing agreement between the two ways of determining this important quantity is found.
However, it needs to be noted that at this point our result from twisted mass lattice QCD has a 
significantly larger error than the one from the dispersive analysis.

Our result also agrees with the value $a_{\mathrm{\mu}}^{\rm hvp} =  6.76\cdot 10^{-8} $ obtained for five flavours with the help of
Dyson-Schwinger equations
in~\cite{Goecke:2011pe}, where the systematic uncertainty of this number has been estimated to be about $ 10 \%$.

\subsection{Systematic effects}

Systematic effects play a very important role in any lattice calculation 
and need to be controlled. We therefore provide in this section a comprehensive 
discussion 
of the various systematic uncertainties appearing in our calculation.

\begin{itemize}
\item {\em Finite-size effects}\\
The systematic uncertainty of finite-size effects appears to be small in our computation. 
The ensembles employed for our result in \Eq{eq:amutot} feature values of 
$3.35 < m_\mathrm{PS}~L < 5.93$. 
Restricting our data to the condition $m_\mathrm{PS}~L >
3.8$ yields a total $$a_{\mathrm{\mu}}^{\rm hvp} = 6.73(25)\cdot 10^{-8}$$ after 
combined continuum and chiral extrapolation which is fully 
compatible with the value quoted in \Eq{eq:amutot}. 

Furthermore, there exist ensembles to explicitly check the volume dependence 
of $a_{\mathrm{\mu}}^{\rm hvp}$ for a pion mass 
of about $320 \mathrm{~MeV}$. For the 
B35 ensembles (see \tab{tab:ensembles}) we have two different
volumes ($32^3 \times 64$ and $48^3 \times 96$) at our disposal. A comparison is given in \tab{tab:FSE}.
\begin{table}[htb]
\begin{center}
\begin{tabular}{|c c| c c|}
\hline
 & & & \vspace{-0.40cm} \\
Ensemble & $\left(\frac{L}{a}\right)^3 \cdot \frac{T}{a}$ & $a_{\mathrm{\mu, ud}}^{\mathrm{hvp}}$& $a_{\mathrm{\mu}}^{\mathrm{hvp}}$ \\
\hline \hline
 & & & \vspace{-0.40cm} \\
B35.32 & $32^3 \times 64$ & $5.45(18)\cdot 10^{-8}$ &   $6.35(19)\cdot 10^{-8}$ \\
B35.48 & $48^3 \times 96$ & $5.46(18) \cdot 10^{-8}$ &  $6.44(19)\cdot 10^{-8}$\\
\hline
\end{tabular}
\caption{ Comparison of light-quark contribution to $a_{\mathrm{\mu}}^{\rm hvp}$ 
and total $a_{\mathrm{\mu}}^{\mathrm{hvp}}$ from
ensembles of different volumes. See \protect\tab{tab:ensembles} for a description 
of the ensembles used here.} 
\label{tab:FSE}
\end{center}
\end{table}

We conclude that finite 
size effects are negligible compared to our statistical error  
and we therefore do not take them as a systematic error into account.

\item {\em Chiral extrapolation} \\
Also the systematic uncertainty of the chiral extrapolation is small. 
The ensembles utilised to arrive at the result given in \Eq{eq:amutot} have values of
$227\,{\rm MeV} < m_\mathrm{PS} < 491\,{\rm MeV}$. 
Restricting our data to the condition $m_\mathrm{PS} < 400 \mathrm{~MeV}$ 
yields the value $$a_{\mathrm{\mu}}^{\rm hvp} = 6.76(26)\cdot
10^{-8}$$ after 
combined continuum and chiral extrapolation which is 
fully compatible with the result quoted in \Eq{eq:amutot}. 
Hence, we do not assign a systematic 
uncertainty of our way of extrapolating to the physical point.

\item {\em Fit ranges of the vector meson fits} \\
The vector meson properties play an important role in our analysis 
of the vacuum polarisation function. In order to estimate the 
systematic effect of determining the vector meson masses and decay constants
we have varied the fit ranges of the single-state vector meson 
fits by a) $0.1\,\mathrm{fm}$ to the left, b) $0.1\,\mathrm{fm}$ to the right, and c)
$0.1\,\mathrm{fm}$ to the left and to the right. 

With this procedure, we do not find  any significant 
differences in the values for $a_{\mathrm{\mu}, s}^{\rm hvp}$ and
$a_{\mathrm{\mu, c}}^{\rm hvp}$. 
For the light quark contribution we find a systematic shift due to
excited state contaminations when including time slices 
corresponding to a shift of $0.1\,\mathrm{fm}$ to 
the left of our standard fit range. This can be seen in
\fig{fig:light_fitrange}.

\begin{figure}[htb]
\centering
\includegraphics[width=0.45\textwidth]{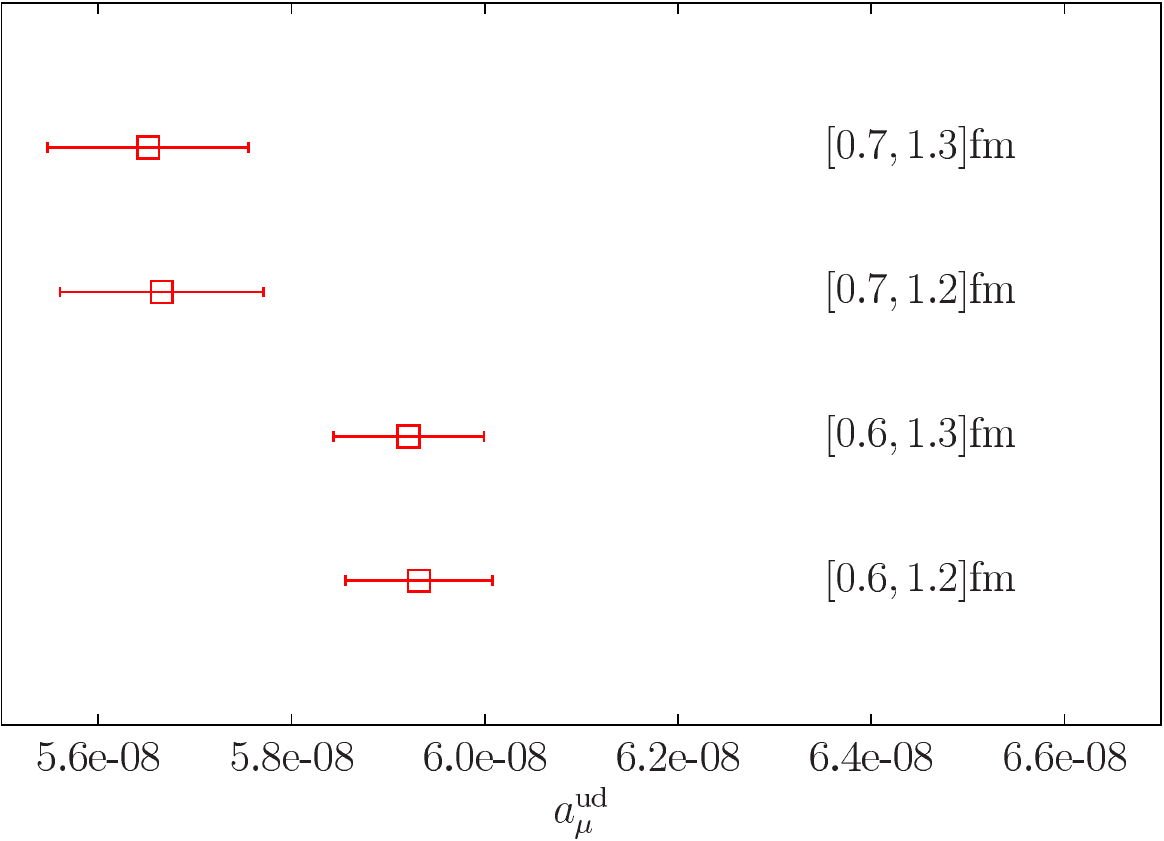}
\caption{Comparison of the effect of choosing different fit ranges on 
$a_{\mathrm{\mu, ud}}^{\rm hvp}$ extrapolated linearly to the
physical
pion mass. The standard fit range is $[0.7\,\mathrm{fm},1.2\,\mathrm{fm}]$.} 
\label{fig:light_fitrange}
\end{figure}

Taking half the difference of $a_{\mathrm{\mu, ud}}^{\rm hvp}$ values 
obtained after fitting the $\rho$-meson in the fit ranges
$[0.6\,\mathrm{fm},1.2\,\mathrm{fm}]$ and $[0.7\,\mathrm{fm},1.2\,\mathrm{fm}]$, 
we obtain a systematic uncertainty from the choice of fit range for
the vector mesons of
\begin{equation}
 \Delta_{V} = 0.13 \cdot 10^{-8} \; .
\end{equation}

\item {\em Fit function} \\
For our MNBC fit functions presented in Sec.~\ref{sec:vacpol} different 
values of $M$,$N$, $B$, and $C$ have been tested with the result that
changing $B$ and $C$ does not have noticeable effects as long as a smooth 
matching between the low-momentum and the high-momentum fit functions is ensured.
For the strange and the charm quark contributions to $a_{\mathrm{\mu}}^{\rm hvp}$ 
even choosing different combinations of $ M \in {1,2}$ and $N \in
{2,3,4}$ does not result in systematic differences. 
However, we do see systematic effects when varying $M$ and $N$ for the light quark contribution.
This is shown in \fig{fig:light_mnbc}.

\begin{figure}[htb]
\centering
\includegraphics[width=0.5\textwidth]{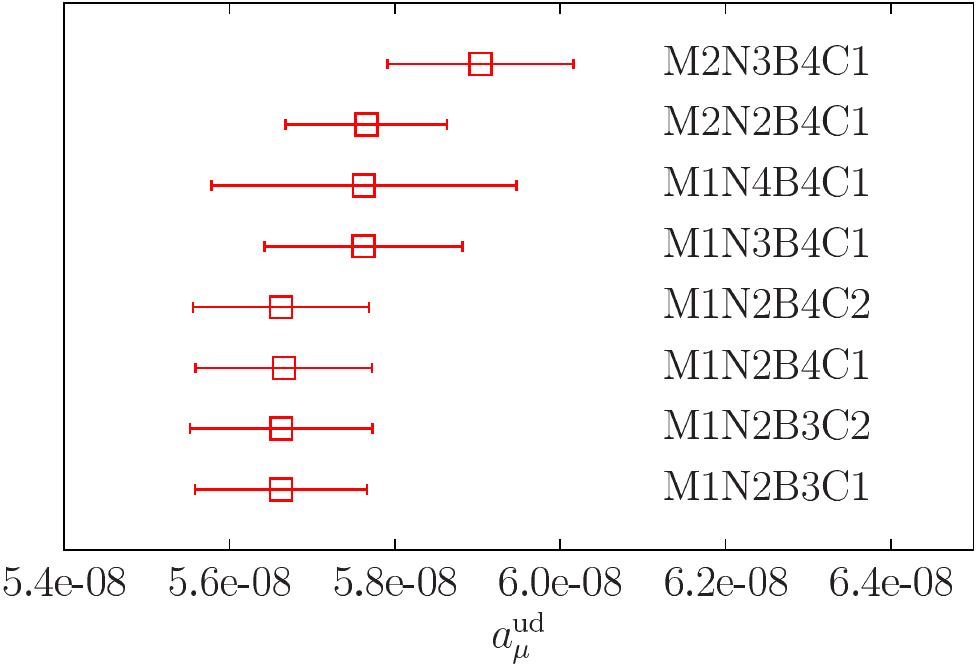}
\caption{Comparison of the effect of different $M$, $N$, $B$, $C$ values 
on $a_{\mathrm{\mu, ud}}^{\rm hvp}$ extrapolated linearly to the physical
pion mass. The standard fit is M1N2B4C1.} 
\label{fig:light_mnbc}
\end{figure}

Thus for the choice of the light quark fit function we take half the difference of the
extrapolated M1N2B4C1 and M2N3B4C1 results as estimate of the systematic uncertainty of choosing values for $M$, $N$, $B$, and $C$:
\begin{equation}
 \Delta_{MNBC} = 0.12 \cdot 10^{-8} \; .
\end{equation}

Additionally, we have checked that varying the matching momentum between 
$1\,\mathrm{GeV}^2$ and $3\,\mathrm{GeV}^2$ gives compatible results for
$a_{\mu}^{\mathrm{hvp}}$ as long as the transition between the fit functions 
in the low- and high-momentum regions is smooth. 
Another criterion 
has been
that the coefficients of the fit polynomial in the high-momentum region, where more data is available,
do not influence the coefficients of the fit polynomial in the low-momentum region.
Applying these criteria all choices of different functions to combine the two momentum regions 
have not resulted in significant differences in the final values for 
$a_{\mathrm{\mu}}^{\rm hvp}$ compared to the
Heaviside step function that we have used for our quoted result.

In~\cite{Aubin:2012me} it was suggested to utilise 
so-called correlated [1,1] Pad\'e approximants to fit the vacuum polarisation function 
in the low momentum region. We tried such fits using an upper integration scale 
of $1.5\,{\rm GeV}^2$. 
Compared to our standard M1N2 fit, with the same upper integration limit, we obtain 
compatible results
for the light $a_{\rm \mu, ud}^{\mathrm{hvp}}$ employing the standard definition of \Eq{eq:amudef}. 
Performing Pad\'e fits for the redefinition in \Eq{eq:redef} using $H=m_V$ determined from the vector meson two-point
function we, however, observe a larger 
error than from our M1N2 fits. This is caused by the fact 
that for the Pad\'e fits the pole is fitted simultaneously and not taken as an external input parameter and thus the
favourable error cancellation between the pole and the vector meson mass which holds for 
the M1N2 fits for the redefinition of \Eq{eq:redef} with $H=m_V$ does not occur. 
In fact, we find that 
fitting the vacuum polarisation function by the Pad\'e ansatz the 
values of $ a_{\rm \overline{\mu}, ud}^{\mathrm{hvp}}$ obtained with the
redefinition show up to three times larger uncertainties.

\item {\em OS matching uncertainties} \\
Since we use the OS action in the strange and charm quark sector, 
different values for the corresponding quark masses could be used which, 
however, have to lead to the physical values of the Kaon and D-meson masses
in the continuum and chiral limit. 
Varying the strange and charm quark masses within the uncertainties given for 
$a \mu_s$ and $a \mu_c$ in Sec.~\ref{sec:setup} has been found
to be negligible. Likewise changing $\mu_s$ to the value obtained from directly matching with the physical kaon mass gives a compatible result. The
same is true when using the $\mu_s$ and $\mu_c$ values procured when allowing for $a^2$-effects in the fit function employed in the matching.

\begin{table}[tb]
\begin{center}
\begin{tabular}{|c| c c c|c|}
\hline
Ensemble & $a_{\mathrm{\mu, ud}}^{\mathrm{hvp}}$& $a_{\mathrm{\mu, s}}^{\mathrm{hvp}}$ & $a_{\mathrm{\mu, c}}^{\mathrm{hvp}}$ &
$a_{\mathrm{\mu}}^{\mathrm{hvp}}$\\
\hline \hline
 & & & & \vspace{-0.40cm} \\
A100.24 & $5.18(18) \cdot 10^{-8}$ & $8.25(59)\cdot 10^{-9}$ & $3.16(24) \cdot 10^{-9}$& $6.32(19)  \cdot 10^{-8} $ \\
A100.24s & $5.32(18) \cdot 10^{-8}$ & $8.87(55)\cdot 10^{-9}$ & $3.16(22) \cdot 10^{-9}$& $6.52(19) \cdot 10^{-8} $\\
\hline
\end{tabular}
\caption{ Comparison of single-flavour contributions
and total $a_{\mathrm{\mu}}^{\mathrm{hvp}}$ from
ensembles having different strange and charm sea quark masses.} 
\label{tab:sea}
\end{center}
\end{table}

\item {\em Different strange and charm sea quark masses}\\
Additionally to the choice of valence quark masses, our result might be influenced by sea quark masses which for
some of the ensembles have not been tuned to their correct physical values. For details see~\cite{Baron:2010bv}. By changing the mass splitting
parameter $\mu_{\delta}$ of the twisted mass action for a non-degenerate fermion doublet \Eq{eq:Sheavy} for the ensemble with the biggest deviation
from the physical strange quark mass, ETMC has generated an ensemble in which both heavy quark masses are compatible with their physical values. The
new ensemble is called A100.24s whereas the old one is A100.24, sharing apart from $\mu_{\delta}$ the same parameters. They have both been tuned to
maximal twist and possess a pion mass of about $500\,\mathrm{MeV}$ and a space-time volume in units of the lattice spacing of $24^3 \times 48$. Due to
the large pion mass they are not included in the rest of our analysis. Using the same matching condition for the OS valence quarks, we arrive at
consistent 
values for the single-flavour contributions to $a_{\mathrm{\mu}}^{\mathrm{hvp}}$ as can be seen in \tab{tab:sea}. Hence, we conclude that the impact of different sea quark masses in the heavy sector on $a_{\mathrm{\mu}}^{\mathrm{hvp}}$ is negligible.

\item {\em Disconnected contributions} \\
This is a systematic effect we can currently not adequately quantify. There are, however, several reasons for assuming that the disconnected
contributions are small. First of all, a dedicated study in the 2-flavour case has revealed them to be compatible with
zero~\cite{Feng:2011zk}. Note, however, that in Refs.~\cite{DellaMorte:2010aq, Francis:2013fzp} the impact of
disconnected contributions has been estimated to be $-10 \%$, at least in the energy range $2 m_{\pi} < q < 3 m_{\pi}$.
Secondly, in the $SU(3)$ flavour limit they are identically zero due to charge cancellation. Thirdly, the disconnected contribution arising from the
charm quark has been computed in perturbation theory and shown to be suppressed by a factor $\left(\frac{q^2}{4~m_c^2}\right)^4$~\cite{Groote:2001py},
where $q^2$ is the relevant energy scale of the problem, i.e. $0.003\,{\rm GeV}^2$.

\end{itemize}

\section{Conclusions}

In this paper, we have performed a first calculation of the four-flavour
leading-order QCD
contribution to the muon anomalous magnetic moment. Such a four-flavour
computation has the invaluable advantage that a direct comparison
to a phenomenological extraction of this quantity can be achieved
without any ambiguity when discriminating the different flavour contributions which appears
unavoidably in a two- or three-flavour comparison.

In our work we found that also for the four-flavour situation
considered here, the improved method of Ref.~\cite{Feng:2011zk} leads to a
smooth and well-controlled chiral extrapolation of $a_{\mathrm{\mu}}^{\rm hvp}$.
In addition to the chiral extrapolation, we performed a comprehensive
analysis of systematic effects, such as finite lattice spacing and finite
volume artefacts, excited state contaminations in the vector meson mass
determination, different choices of the vacuum polarisation fitting
function, and different choices of valence and sea strange and charm quark masses.
From this set of systematic effects only the different choices of the
fitting functions and excited state contaminations of the vector states lead to significant
systematic uncertainties which we include in our final error estimate.

\begin{figure}[htb]
\centering
\includegraphics[width=0.45\textwidth]{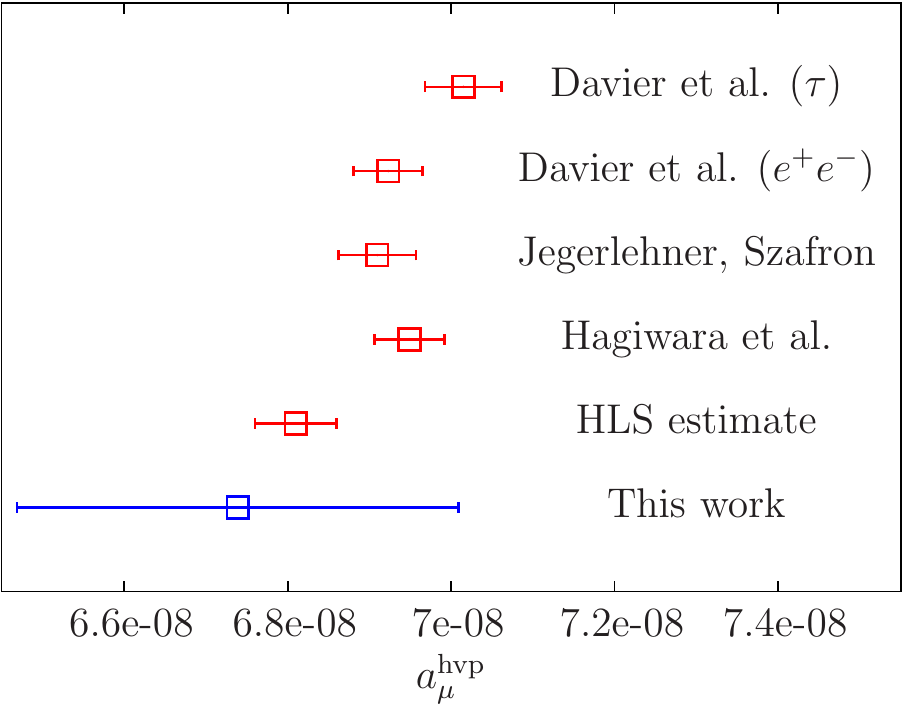}
\caption{Comparison of our first four-flavour lattice result
of $a_{\mathrm{\mu}}^{\rm hvp}$ with different results based on dispersion relations:
Davier et al.~\protect\cite{Davier:2010nc},
Jegerlehner and Szafron~\protect\cite{Jegerlehner:2011ti},
Hagiwara et al.~\protect\cite{Hagiwara:2011af}, and HLS~\protect\cite{Benayoun:2012wc}}
\label{fig:amutot_comp}
\end{figure}

As our main result we provide a comparison to
a dispersive analysis~\cite{Jegerlehner:2008zz}:
\begin{eqnarray}
a_{\mathrm{\mu}}^{\rm hvp} & = & 6.74(21)(18)\cdot 10^{-8}\;\;\; (N_f=2+1+1)  \nonumber\\
a_{\mathrm{\mu}}^{\rm hvp} & = & 6.91(01)(05)\cdot 10^{-8}\;\;\; (\mathrm{dispersive}\; \mathrm{analysis})\;. 
\end{eqnarray}
In \Fig{fig:amutot_comp} we also compare the outcome of our first-principle 
computation with a summary of other results obtained utilising the dispersion relation.
Although our lattice QCD determination of $a_{\mathrm{\mu}}^{\rm hvp}$
shows an overall agreement
with phenomenology,
the lattice QCD result has clearly a
significantly larger error being, however, already  at the same order of magnitude.

A substantial step forward to improve the lattice determination
will be computations directly at the physical value of the pion mass.
Such simulations, in combination with a significantly increased statistics including also isospin breaking and
electromagnetic effects
have the potential to reach or even beat the error from a dispersive
analysis. In addition, results from lattice QCD will only rely on QCD alone
and hence can provide a stringent test of the standard model.
This opens
the exciting possibility to eventually clarify whether
the present discrepancy for the muon anomalous magnetic moment is indeed
a sign of new physics.

\label{sec:conc}

\section*{Acknowledgements}
We thank the European Twisted Mass Collaboration (ETMC) for generating the gauge field ensembles used in this work and Andreas Ammon for
providing us with the information of the matching K- and D-meson 
masses in the mixed-action setup with their physical values. Special thanks goes to Elena Garcia-Ramos and Krzysztof Cichy for enlightening
discussions concerning the $\mathcal{O}(a)$ improvement.
This work has been supported in part by the DFG Corroborative
Research Center SFB/TR9.
G.H.~gratefully acknowledges the support of the German Academic National Foundation (Studienstiftung des deutschen Volkes e.V.) and of the
DFG-funded Graduate School GK 1504.
K.J. was supported in part by the Cyprus Research Promotion
Foundation under contract $\Pi$PO$\Sigma$E$\Lambda$KY$\Sigma$H/EM$\Pi$EIPO$\Sigma$/0311/16.
This manuscript has been coauthored by Jefferson Science Associates, LLC under Contract No.~DE-AC05-06OR23177 with the U.S.~Department of Energy.
The numerical computations have been performed on the
{\it SGI system HLRN-II} at the {HLRN Supercomputing Service Berlin-Hannover},  FZJ/GCS, BG/P, and BG/Q at FZ-J\"ulich.

%\newpage
\appendix
\section{Data}
\begin{table}[htb]
\begin{center}
\begin{tabular}{|c| c c |c| c c |c|}
\hline
 & & & & & & \vspace{-0.40cm} \\
Ensemble & $f_\rho$[GeV] & $m_\rho$[GeV] & $\chi^2/dof$ & $f_{s\overline{s}}$[GeV] & $m_{s\overline{s}}$[GeV] & $\chi^2/dof$\\
 & & & & & &\vspace{-0.40cm} \\
\hline \hline
& & & & & & \vspace{-0.40cm} \\
D15.48 &  0.263(17) & 0.933(55) & 1.01 & 0.289(08) & 1.129(14) & 1.06 \\
D30.48 &  0.279(12) & 0.998(34) & 0.21 & 0.291(10) & 1.132(16) & 0.79  \\
D45.32sc& 0.279(09) & 1.003(27) & 0.52 & 0.273(06) & 1.112(13) & 1.06  \\
% & & & & \vspace{-0.40cm} \\
%
\hline
 & & & & & &\vspace{-0.40cm} \\
B25.32t& 0.261(14) & 0.903(44) & 1.23 & 0.268(12) & 1.112(21) & 0.88 \\
B35.32 & 0.269(15) & 0.967(47) & 0.47 & 0.271(09) & 1.129(19) & 1.28 \\
B35.48 & 0.288(16) & 1.028(44) & 1.13 & 0.286(18) & 1.156(35) & 1.77 \\
B55.32 & 0.278(11) & 0.996(33) & 1.81 & 0.294(10) & 1.144(17) & 0.10 \\
B75.32 & 0.289(10) & 1.056(28) & 1.02 & 0.285(15) & 1.154(26) & 0.15 \\

\hline
 & & & & & &\vspace{-0.40cm} \\

%
%$1.90$ & $0.086$ & $24^3 \times 48$ & $0.0100$ & $252$      \\
%
A30.32 & 0.264(25) & 0.954(68) & 1.18 &  0.284(14) & 1.149(27) & 1.16\\
A40.32 & 0.238(13) & 0.863(39) & 0.81 &  0.278(12) & 1.135(20) & 1.55\\
A50.32 & 0.271(16) & 0.992(42) & 1.50 &  0.268(12) & 1.117(22) & 1.08 \\
\hline
\end{tabular}
\caption{ $\rho$- and $\phi$-meson masses and decay constants obtained from correlator fits outlined in
Sec.~\protect\ref{sec:vectormesons}.} 
\label{tab:paras_rho_ss}
\end{center}
\end{table}

\begin{table}[htb]
\begin{center}
\begin{tabular}{|c| c c |c|}
\hline
 & & & \vspace{-0.40cm} \\
Ensemble & $f_{J/\Psi}$[GeV] & $m_{J/\Psi}$[GeV] & $\chi^2/dof$ \\
 & & & \vspace{-0.40cm} \\
\hline \hline
& & &  \vspace{-0.40cm} \\
D15.48 & 0.439(07) & 3.079(05) & 0.99\\
D30.48 & 0.435(06) & 3.078(04) & 0.44 \\
D45.32sc&  0.429(05) & 3.070(03) & 0.83  \\
% & & & & \vspace{-0.40cm} \\
%
\hline
 & & & \vspace{-0.40cm} \\
B25.32t& 0.443(08) & 3.057(05) & 0.66 \\
B35.32 & 0.437(08) & 3.051(05) & 0.95 \\
B35.48 & 0.412(11) & 3.035(08) & 1.45 \\
B55.32 & 0.446(09) & 3.053(06) & 1.63 \\
B75.32 & 0.447(12) & 3.065(08) & 0.93 \\

\hline
 & & & \vspace{-0.40cm} \\

%
%$1.90$ & $0.086$ & $24^3 \times 48$ & $0.0100$ & $252$      \\
%
A30.32 & 0.438(09) & 3.039(06) & 2.42 \\
A40.32 & 0.436(10) & 3.041(06) & 0.37 \\
A50.32 & 0.453(09) & 3.051(06) & 1.50 \\
\hline
\end{tabular}
\caption{ $J/\Psi$ masses and decay constants obtained from correlator fits outlined in Sec.~\protect\ref{sec:vectormesons}.}
 \label{tab:paras_JPsi}
\end{center}
\end{table}

\begin{table}[htb]
\begin{center}
\begin{tabular}{|c| c c c| c |}
\hline 
 & & & &  \vspace{-0.40cm} \\
Ensemble & $a_{\mathrm{\mu, ud}}^{\mathrm{hvp}}$& $a_{\mathrm{\mu, s}}^{\mathrm{hvp}}$ & $a_{\mathrm{\mu, c}}^{\mathrm{hvp}}$ &
$a_{\mathrm{\mu}}^{\mathrm{hvp}}$\\
\hline \hline
D15.48 &  $5.59(16) \cdot 10^{-8}$ &  $6.68(75) \cdot 10^{-9}$ & $2.13(25) \cdot 10^{-9}$ & $6.47(18)\cdot 10^{-8}$\\
D30.48 &  $5.51(15) \cdot 10^{-8}$ &  $7.57(51) \cdot 10^{-9}$ & $2.40(16)  \cdot 10^{-9}$ & $6.51(16)\cdot 10^{-8}$\\
D45.32sc & $5.41(10) \cdot 10^{-8}$ & $7.36(40) \cdot 10^{-9}$ & $2.43(13)  \cdot 10^{-9}$ &  $6.39(11) \cdot 10^{-8}$\\        
\hline
 & & & & \vspace{-0.40cm} \\
B25.32t&  $5.70(60) \cdot 10^{-8}$ & $5.96(56)\cdot 10^{-9}$ & $2.12(21)\cdot 10^{-9}$ &  $6.50(17)\cdot 10^{-8}$\\
B35.32 & $5.45(18)\cdot 10^{-8}$ &   $6.55(62)\cdot 10^{-9}$ &  $2.41(23)\cdot 10^{-9}$ & $6.35(19)\cdot 10^{-8}$\\
B35.48 & $5.46(18) \cdot 10^{-8}$ &  $7.23(63)\cdot 10^{-9}$ & $2.61(23)\cdot 10^{-9}$ & $6.44(19)\cdot 10^{-8}$\\
B55.32 & $5.44(14) \cdot 10^{-8}$ &  $7.40(44)\cdot 10^{-9}$ & $2.53(17)\cdot 10^{-9}$ &  $6.43(15)\cdot 10^{-8}$\\ 
B75.32 & $5.23(12) \cdot 10^{-8}$ &  $7.67(45)\cdot 10^{-9}$ & $2.79(16)\cdot 10^{-9}$ &  $6.28(13)\cdot 10^{-8}$\\
B85.24 & $5.15(16)\cdot 10^{-8}$ &   $8.21(51)\cdot 10^{-9}$ & $3.11(18)\cdot 10^{-9}$ &  $6.28(17)\cdot 10^{-8}$\\
\hline
 & & & &  \vspace{-0.40cm} \\
A30.32 & $5.40(28)\cdot 10^{-8}$ &  $6.42(89)\cdot 10^{-9}$ &  $2.34(34) \cdot 10^{-9}$ &  $6.28(30)\cdot 10^{-8}$\\
A40.32 & $5.25(17) \cdot 10^{-8}$ &  $5.33(48) \cdot 10^{-9}$ &  $1.92(18)\cdot 10^{-9}$ & $5.98(18)\cdot 10^{-8}$\\
A50.32 & $5.28(20) \cdot 10^{-8}$ &  $6.98(58)  \cdot 10^{-9}$ & $2.58(22) \cdot 10^{-9}$ & $6.23(21)\cdot 10^{-8}$ \\
\hline
\end{tabular}
\caption{ Values for single-flavour contributions as well as total $a_{\mathrm{\mu}}^{\rm hvp}$ obtained in our calculation.}
\label{tab:amudata}
\end{center}
\end{table}

\clearpage
\bibliographystyle{JHEP}
\bibliography{amu_letter}

\end{document}